\documentclass[iop, twocolumn, apj]{emulateapj}

\usepackage{Sweave}
\usepackage[OT1]{fontenc}

\usepackage[pdftex,pdfpagemode={UseOutlines},bookmarks,bookmarksopen,colorlinks,linkcolor={blue},citecolor={black},urlcolor={red}]{hyperref}
\usepackage{graphicx}
\usepackage{bm}
\usepackage{amsmath,amssymb}
\usepackage{aas_macros}

\usepackage{longtable}
\usepackage{booktabs}

\newcommand{\sobs}{\hat{S}} 
\newcommand{\sedparams}{\boldsymbol \theta} 

\newcommand{\kf}{k_{f}}
\newcommand{\ks}{k_{s}}
\newcommand{\nf}{n_{0,f}}
\newcommand{\nsa}{n_{0,s}}
\newcommand{\nsb}{n_{1,s}}
\newcommand{\sflat}{S_{0,f}}
\newcommand{\ssteep}{S_{0,s}}
\newcommand{\lf}{\ell_{f}}
\newcommand{\ls}{\ell_{s}}

\newcommand{\lsed}{\mathit{l}}

\newcommand{\smin}{S_{\rm min}}

\newcommand{\scut}{S_{\rm cut}}

\slugcomment{LLNL-JRNL-501091}
\shorttitle{CMB faint inverted radio foregrounds}
\shortauthors{Schneider et al.}

\begin{document}

\setkeys{Gin}{width=0.5\textwidth}
  
\title{FOREGROUND PREDICTIONS FOR THE COSMIC MICROWAVE BACKGROUND POWER SPECTRUM FROM
MEASUREMENTS OF FAINT INVERTED RADIO SOURCES AT 5 GHz}

\author{Michael D. Schneider}
\affil{Lawrence Livermore National Laboratory, P.O. Box 808 L-210, Livermore, CA 94551, USA}
\email{schneider42@llnl.gov}
\author{Robert H. Becker}
\affil{Department of Physics, University of California, One Shields Avenue, Davis, CA 08991, USA \\
Lawrence Livermore National Laboratory, P.O. Box 808 L-415, Livermore, CA 94551, USA}
\author{Willem de Vries}
\affil{Lawrence Livermore National Laboratory, P.O. Box 808 L-211, Livermore, CA 94551, USA}
\author{Richard L. White}
\affil{Space Telescope Science Institute, Baltimore, MD 21218, USA}

\begin{abstract}
  We present measurements of a population of matched radio sources at 1.4 and 5~GHz 
  down to a flux limit of 1.5~mJy in 7~deg$^{2}$ of the NOAO Deep Field South.  
  We find a significant fraction of sources with inverted spectral indices that 
  all have 1.4~GHz fluxes less than 10~mJy and are therefore too faint to have been 
  detected and included in previous radio source count models that are matched at 
  multiple frequencies.
  Combined with the matched source population at 1.4 and 5~GHz in 1~deg$^{2}$ in the 
  ATESP survey, we update models for the 5~GHz differential number counts and distributions 
  of spectral indices in 5~GHz flux bins that can be used to estimate the unresolved 
  point source contribution to the cosmic microwave background temperature anisotropies.
  We find a shallower logarithmic slope in the 5~GHz differential counts than in previously 
  published models for fluxes $\lesssim 100$~mJy as well as larger fractions of inverted 
  spectral indices at these fluxes. 
  Because the Planck flux limit for resolved sources is larger than 100~mJy in all channels, 
  our modified number counts yield at most a 10\%
  change in the predicted Poisson contribution to the Planck temperature power spectrum.
  For a flux cut 
  of 5~mJy with the South Pole Telescope and a flux cut of 20~mJy with the Atacama Cosmology Telescope
  we predict a $\sim$30\%  and 
  $\sim$10\%
  increase, respectively, in the radio source Poisson power in the lowest frequency 
  channels of each experiment relative to that predicted by previous models.

\end{abstract}


\maketitle

\section{Introduction}
\label{sec:introduction}
A main science driver for ongoing high-resolution cosmic microwave background (CMB) 
experiments is to measure the amplitude and tilt of the power spectrum of 
primordial density fluctuations and thereby constrain models of cosmic inflation.
On large angular scales (multipoles $\ell \lesssim 1000$), the power spectrum of 
CMB temperature anisotropies is directly related to the primordial 
density power spectrum by well-understood baryon physics.  
However, on smaller angular scales various astrophysical foregrounds dominate the 
measured temperature power spectrum and obscure the primordial signal.
  
For frequencies less than about 150~GHz a dominant 
foreground is the flux from radio-loud galaxies and blazars that is unresolved by the  
CMB experiment beam and therefore cannot be masked out \citep{1996MNRAS.281.1297T,1999MNRAS.307..977K,1999AA...346....1S,2008ApJ...688....1H,2010MNRAS.407..247C,2011arXiv1102.5195M}. 
Previous radio surveys have identified a significant population of sources from 1~to~30~GHz that have 
flat or ``inverted'' spectral energy distributions (SEDs) such that their flux is roughly constant or 
increasing with increasing frequency~\citep{Guerra:2002im,2010MNRAS.407..247C, Prandoni:2010gv} 
and are therefore potentially significant foregrounds for CMB observations.  
Physical models for these objects predict that they are dominated by synchrotron emission from 
relativistic jets in active galactic nuclei~\citep[AGNs;][]{1987ApJ...318L..15D,1998MNRAS.297..117T,2005A&A...431..893D,2010A&ARv..18....1D,2011A&A...533A..57T, 2011MNRAS.417.1881R} with the large spectral indices determined either by dominant emission 
from compact optically thick regions of the jet, early or late stages of AGN evolution, 
or chance variability in the source. Such physical models predict that the SEDs of these sources should break and decline somewhere between tens and hundreds of GHz.
The flat and inverted radio sources detected around a few GHz should then contribute negligible 
flux at millimeter frequency CMB observations (where infrared emission from 
galactic dust becomes a problem instead).

Thermal synchrotron emission from advection-dominated accretion flows 
(ADAFs) may be also be a significant contribution to the low-frequency radio foregrounds in CMB experiments~\citep{2000ApJ...542...68P, 2004MNRAS.354.1005P, 2005A&A...438..475T}. ADAFs may also show inverted spectra up to several tens of GHz when the emission region is
significantly compact. Using the model of~\citet{2000ApJ...542...68P} and 
considering constraints on the source counts from the Ryle telescope, ATCA, and 
{\it Wilkinson Microwave Anisotropy Probe (WMAP)} surveys, \citet{2005A&A...431..893D} showed that the predicted differential 
number counts of ADAFs at 30~GHz are at least two orders of magnitude smaller 
than the blazar counts for fluxes greater than $\sim 1$~mJy (see their Figure 14).
We therefore only consider SED models for blazars with inverted spectra and 
neglect any possible (sub-dominant) contribution from ADAF sources.

Aside from CMB power spectrum measurements, radio foregrounds are also a significant foreground 
for Sunyaev--Zel'dovich measurements of galaxy clusters~\citep{2004ApJ...612...96K,2004ApJ...602..565W, 2005A&A...431..893D, 2009ApJ...694..992L, 2010ApJ...709..920S, 2011arXiv1108.3343R} 
and for testing for non-Gaussianity in the CMB~\citep{2010A&A...513A..59E,2011MNRAS.417..488C}.

As CMB experiments obtain better resolution, the lower flux limit will decrease for resolving and 
removing radio sources.  It is then important to characterize the radio source number counts and 
spectral indices at fluxes below the resolved source flux cut to mitigate the contamination 
from unresolved sources.
While {\it WMAP}\footnote{\url{http://map.gsfc.nasa.gov}} 
was able to remove radio point sources with fluxes greater than 0.7~Jy,  
Planck\footnote{\url{http://www.esa.int/planck}} resolves all sources with fluxes 
greater than~$\sim0.2$~Jy~\citep{PlanckERCSC}.  
The Atacama Cosmology Telescope (ACT)\footnote{\url{http://www.physics.princeton.edu/act/}}
and the South Pole Telescope (SPT)\footnote{\url{http://pole.uchicago.edu}} are able 
to resolve all sources above 20 and 5 mJy respectively~\citep{2011ApJ...731..100M, 2010ApJ...719..763V}.
Existing models for radio source counts and SEDs have largely been calibrated only to a flux limit of 
30-100~mJy~\citep{2005A&A...431..893D,2011A&A...533A..57T}, so the models would have to be extrapolated 
to lower fluxes in order to analyze these ongoing CMB experiments.

We present measurements of the population of faint radio sources complete to 1.5~mJy at 1.4 and 5~GHz 
collected with the Very Large Array (VLA) in 7~deg$^{2}$ of the NOAO Deep Field South (DFS).  Our sources 
are matched at the two observation frequencies allowing us to determine the spectral index 
for each source.
To apply our source catalog for predicting the unresolved point source contamination in the CMB, we compare primarily with~\citet{2011A&A...533A..57T} (hereafter T11) who present models for the 
number counts and SEDs of radio sources calibrated from many different data sets.  
Our data are new primarily in reaching fainter flux densities with spectral indices of matched sources.

This paper is organized as follows.
We describe our observations and the construction of our source catalog in 
Section~\ref{sec:catalog}. We derive new fits to the 5~GHz differential number counts 
of our sources and the distributions of spectral indices in Section~\ref{sec:countmodel}. 
We then review SED models that use the 
fits to the 5~GHz counts and indices as inputs for 
extrapolating source fluxes to higher frequencies in Section~\ref{sec:extrapolation}.
We show the impact of our new data and models on the predicted high-frequency number 
counts in Section~\ref{sec:numbercounts} and predict the impact on measurements of the 
Planck temperature--temperature (TT) power spectrum in Section~\ref{sec:cmb}
We draw conclusions about the impact of faint radio sources on future CMB measurements 
in Section~\ref{sec:conclusions}.  
In Appendix~\ref{sec:errorpropagation} we describe a statistical 
model for propagating the errors in the measured fluxes and spectral indices into the 
estimated high-frequency differential counts and CMB Poisson contribution from 
unresolved sources. We describe the structure of our new source catalog in 
Appendix~\ref{sec:catalogtable} and present a sample of the measured source fluxes in Table~\ref{tab:noaocat}. The full catalog is available for download from the VizieR database.


\section{Source catalog}
\label{sec:catalog}
We present a catalog of 362 new sources discovered 
with VLA imaging in 
7~deg$^2$ of the NOAO DFS at 1.4 and 5~GHz (see Appendix~\ref{sec:catalogtable} for a sample of the catalog).  
The data reduction and catalog construction were performed with the same methods as 
in the Faint Images of the Radio Sky at Twenty cm (FIRST) 
survey~\citep{1997ApJ...475..479W}.  The survey has a resolution 
of 5 arcsec at both frequencies.  We include in our catalog only sources 
that are matched in the two bands. There are 700 identified sources in the 
field at 1.4~GHz, so roughly half of the sources are matched at 5~GHz to give 
us our 362 sources. There are seven point sources in our 5~GHz catalog 
that are not matched at 1.4~GHz. All seven sources have 5~GHz fluxes between 1 and 5~mJy, and therefore 
likely have inverted spectral indices such that the 1.4~GHz flux is below the FIRST survey detection threshold 
around 1~mJy. Lacking spectral index measurements for these seven sources, we do not include them 
in our subsequent analysis. However, all our results must be interpreted remembering that there are 
likely more inverted sources near our flux limit that would further boost the differential number 
counts of faint point sources.

The ATESP~\citep{2000A&AS..146...41P,Prandoni:2006rm}
catalog has a similar resolution to ours and has 118 matched sources in the same 
frequency bands over 1~deg$^2$.  The ATESP catalog is complete below 1~mJy and is 
therefore a useful complement to our catalog.  We will consider both the NOAO DFS and 
ATESP catalogs in tandem when fitting models for the differential number counts and 
SEDs of the 5~GHz source population.

\begin{figure}
\begin{center}
\includegraphics[width=0.5\textwidth]{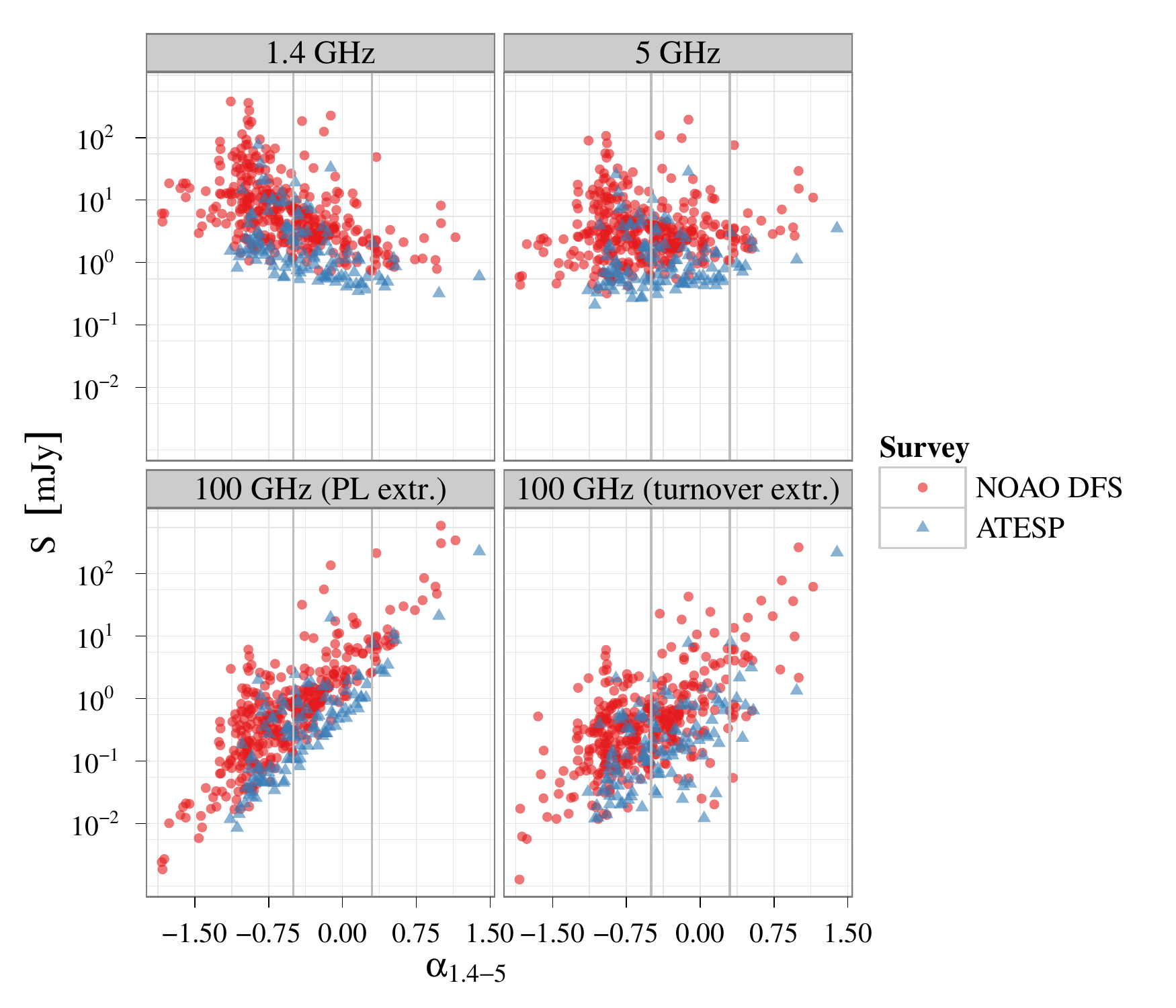}
\caption{\label{fg:fluxvsindex} Flux in the observation bands of 1.4 and 5 GHz vs. the 
spectral index inferred from these bands.  The bottom panels show the fluxes 
extrapolated to 100~GHz using two different SED models as described in 
Section~\ref{sec:extrapolation}.  The bottom left panel uses a simple power-law (PL) model 
while the bottom right panel uses a model from T11 that has a ``turnover'' in the SED 
at $\sim10--30$~GHz.}
\end{center}
\end{figure}

The observed fluxes and the inferred spectral indices 
(defined by the relation $S \propto \nu^{\alpha}$ where $S$ is the flux, $\nu$ is the frequency, and $\alpha$ is the spectral index) 
for each of our matched sources are shown in Figure~\ref{fg:fluxvsindex}.  
There are 28 
and 9 sources with spectral indices 
$\alpha_{1.4-5}>0.3$ in the NOAO~DFS and ATESP catalogs respectively, 
which we label as ``inverted'' spectrum sources (following a common convention). 
All of our inverted spectrum sources have 1.4~GHz fluxes $<10$~mJy so they would 
not have been identified in earlier surveys with larger flux limits.
The sources with $-0.5 <\alpha_{1.4-5}\le0.3$ 
(129 in 
the NOAO DFS catalog and 
55 in the ATESP catalog)
are labeled as ``flat-spectrum'' 
and could also potentially be visible at higher frequencies.  Again, we see in 
the top panels of Figure~\ref{fg:fluxvsindex} that the majority of our flat-spectrum 
sources have 1.4~GHz fluxes below 10~mJy.
As a preliminary illustration of the potential contribution of these source populations for 
high-frequency number counts we show the extrapolated fluxes of our sources at 100~GHz 
using two different SED models in the bottom panels of Figure~\ref{fg:fluxvsindex}.
We will explain the SED models in the next section, but note here that with both a 
``power-law'' (PL) SED model and an SED model with a break or ``turnover'' many of our 
sources are predicted to have 100~GHz fluxes greater than 1~mJy and will therefore be 
detectable in high-resolution CMB experiments.
In our fits to the 5~GHz differential number counts in the next section, it will also be 
important to note that there is not a strong correlation between 
the 5~GHz flux and $\alpha_{1.4-5}$ in Figure~\ref{fg:fluxvsindex}.

\begin{figure}
\begin{center}
\includegraphics[width=0.5\textwidth]{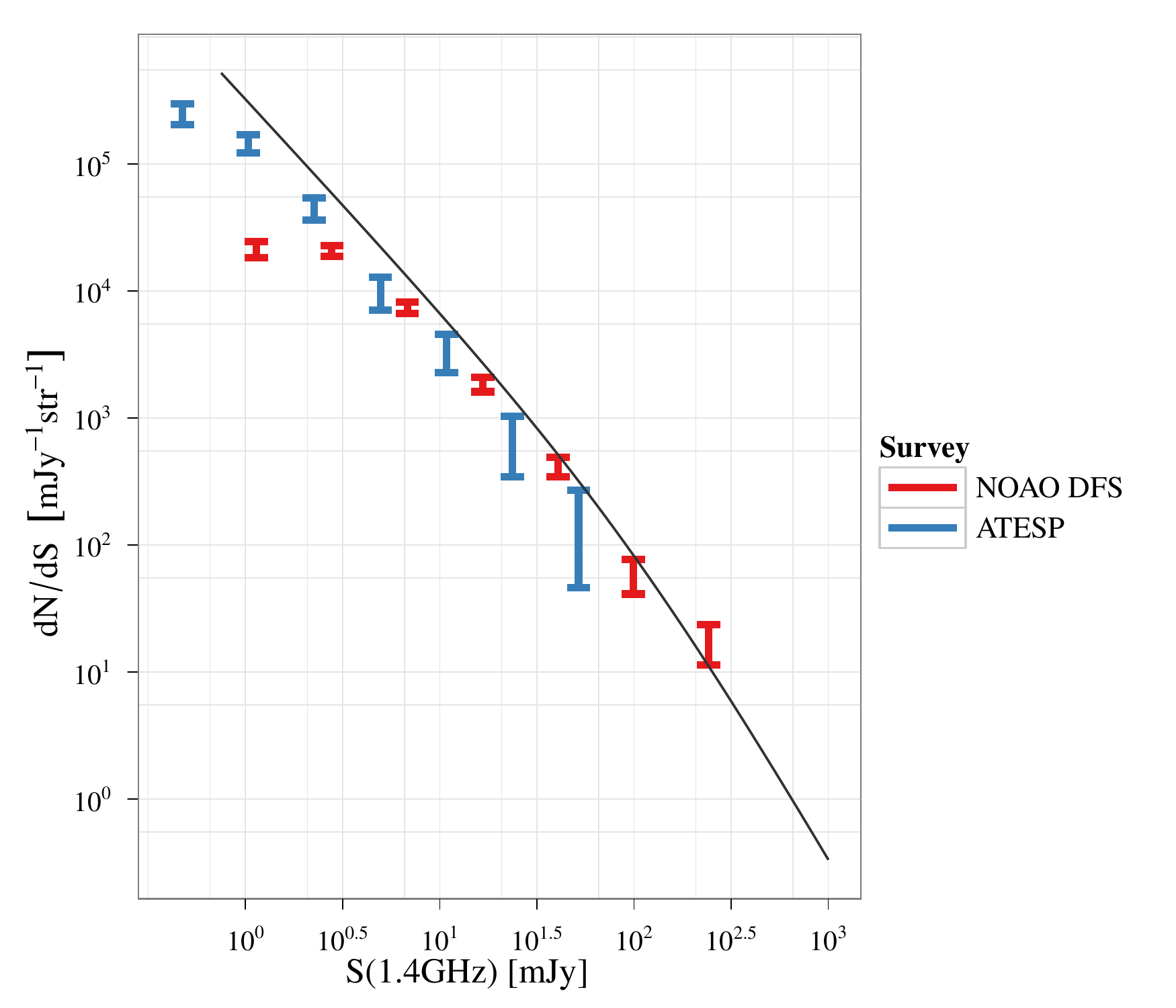}
\caption{\label{fg:dNdS1}Estimated differential number counts at 1.4~GHz.  The solid line is the fit to the 
counts from the VLA~FIRST survey at 1.5~GHz from~\citet{1996MNRAS.281.1297T}, which is complete to 0.75~mJy.}
\end{center}
\end{figure}

In Figure~\ref{fg:dNdS1} we show the differential number counts at 1.4~GHz estimated from 
the NOAO DFS and ATESP catalogs and compare
with the fit to the number counts published by~\citet{1996MNRAS.281.1297T} from the FIRST
survey at 1.4~GHz~\citep{1997ApJ...475..479W}, which is claimed to be complete to 
$S=0.75$~mJy.  Both the NOAO~DFS and ATESP data match the fit well for $S\gtrsim 10$~mJy but the 
NOAO~DFS data 
clearly give fewer number counts for $S\lesssim 3$~mJy while the ATESP counts drop below the fit for 
$S\lesssim 1$~mJy.  
The incompleteness at low flux in both catalogs is primarily caused by the requirement to 
match sources at 5~GHz, which omits steep-spectrum-sources at 1.4~GHz that are not detected at 5~GHz.
This selection effect is apparent in the top left panel of Figure~\ref{fg:fluxvsindex} where the 
minimum flux for steep-spectrum sources ($\alpha_{1.4-5}<-0.5$) is much larger than that for flat$+$inverted 
sources.

\section{Models for source counts and spectral energy distributions}
\label{sec:models}
In this section, we describe the modifications based on our data to the T11 models for 5~GHz number counts, 
spectral index distributions, and SEDs used for extrapolation of the counts to higher frequencies.
The key changes to the T11 model are from the extension to lower 5~GHz fluxes.

\subsection{Models for 5~GHz Sources}
\label{sec:countmodel}

We adopt the form of the fit function from Equation~(1) of~T11 for the 5~GHz 
differential number counts for flat$+$inverted spectrum sources,
\begin{equation}\label{eq:Tucci1}
  \frac{dN_{\rm fl+inv}}{dS_{5}} = \nf \frac{\left(S/\sflat\right)^{\kf}}{1-e^{-1}}
  \left(1 - e^{-\left(S/\sflat\right)^{\lf-\kf}}\right).
\end{equation}
T11 found best fit parameters $\nf=47.4$~Jy$^{-1}$sr$^{-1}$, $\sflat=1.67$~Jy, 
$\kf=0.50$ and $\lf=-0.66$. Because of the small areas covered by the NOAO DFS and ATESP 
catalogs, we do not have good statistics for constraining the differential number 
counts at fluxes greater than about 100~mJy.  
We therefore include the differential count estimates (covering $0.1 \lesssim S \lesssim 10$~Jy) 
presented in the right panel of Figure~2 of T11 
when deriving new best fit parameters for the fitting function in Equation~(\ref{eq:Tucci1}).
In~T11, the steep-spectrum number counts are defined as the 
difference between the $dN/dS_5$ model of~\citet{1998MNRAS.297..117T} 
and Equation~(\ref{eq:Tucci1}).
Here, we introduce a fit function for the steep-spectrum counts that can be used to 
fit our steep-spectrum sources,
\begin{equation}\label{eq:SteepCounts}
  \frac{dN_{\rm st}}{dS_{5}} = \nsa \frac{\left(S/\ssteep\right)^{\ks}}{1-e^{-1}}
  \left(1 - e^{-\left(S/\ssteep\right)^{\ls-\ks}}\right) + 
  \nsb \left(\frac{S}{1\,{\rm Jy}}\right)^{a_1},
\end{equation}
where $\nsa=88$~Jy$^{-1}$sr$^{-1}$, $\ssteep=0.12$~Jy, $\ks=0.84$, $\lf=-0.49$, 
$\nsb=10.9$~Jy$^{-1}$sr$^{-1}$, and $a_1=0.33$ fit the steep-spectrum count model reported in~T11.

When fitting to the differential counts estimated from our catalogs, we multiply 
Equations~(\ref{eq:Tucci1}) and (\ref{eq:SteepCounts}) by selection functions of the form, 
\begin{equation}\label{eq:selectionfunction}
  \phi(S_5; \smin, a) \equiv 1 - e^{-\left(S_5 / \smin\right)^a},
\end{equation}
to account for incomplete source extraction near the limiting fluxes in each catalog.
We fit separate values of $\smin$ and $a$ for the NOAO~DFS and ATESP catalogs. For each catalog, 
we also fit separate $\smin$ and $a$ values for the flat$+$inverted and steep-spectrum samples 
because the flux limit of matched sources is a function of the source spectral index (see, e.g., the skewed lower bounds of the scatter plots in the top panels of Figure~\ref{fg:fluxvsindex}).

We fit new values for the flat+inverted differential number count model in Equation~(\ref{eq:Tucci1}) 
with the following procedure.
\begin{enumerate}
  \item We create separate histograms 
  for the 5~GHz fluxes in the NOAO~DFS and ATESP catalogs with the optimal
  histogram bins chosen as prescribed by~\citet{2008arXiv0807.4820H} 
  (assuming the logarithm of the bin widths is constant).
  The histogram bin values with Poisson uncertainties are our estimators for the 5~GHz differential 
  number counts.
  \item We combine our number count estimates with those presented in the right panel of Figure~2 in T11, 
  which compares the 5~GHz count estimates from a number of surveys covering the flux range 
  $\sim100$~mJy to $\sim10$~Jy. Because our data cover a small area, we do not have good estimates of the 
  number counts above $\sim100$~mJy so the data from T11 is essential to constrain the ``bend'' 
  and high-flux slope of the fit function in Equation~(\ref{eq:Tucci1}). Note that we assign new Poisson 
  uncertainties to the data points extracted from Figure~2 in T11 according to the number of sources in each 
  flux bin (derived according to the area of each survey). The uncertainties assigned to our differential count 
  estimates and those from T11 are therefore consistent.
  \item We extract samples from the joint posterior of all the fit parameters in Equation~(\ref{eq:Tucci1}) and 
  $\smin$ and $a$ (from Equation~(\ref{eq:selectionfunction})) for each of the NOAO~DFS and ATESP data sets using 
  Markov Chain Monte Carlo (MCMC) with a likelihood given by the product of the Poisson likelihoods for each 
  count estimate in each flux bin for each data set (our two catalogs and those from Figure~2 in T11). 
  We thin the MCMC samples to obtain uncorrelated samples of the joint parameter posterior.
  \item We define the best fit parameters as the medians of the marginal posteriors for each parameter.
\end{enumerate}
\begin{figure*}
\centerline{
\includegraphics[width=0.7\textwidth]{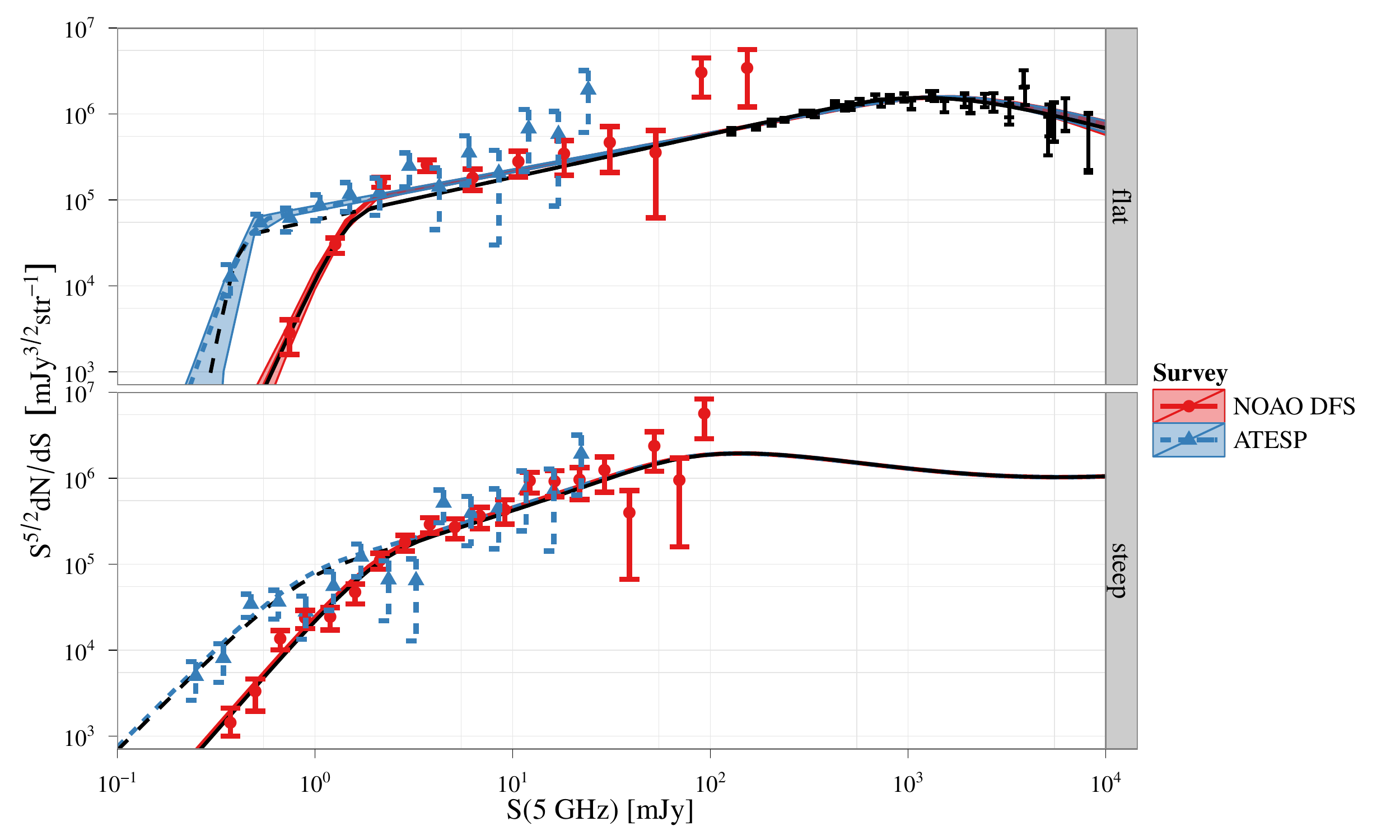}
}
\caption{\label{fig:flatfit}Observed differential number counts at 5~GHz and 
fits for the flat$+$inverted spectrum (top) and steep-spectrum (bottom) sources.
The red solid lines in each panel are the fits to the NOAO~DFS counts (red points and error bars) while the blue dashed lines are the fits to the ATESP counts (blue triangles and error bars).
The solid and long-dashed black 
lines show the fits from~\citet{2011A&A...533A..57T} for the NOAO~DFS and ATESP selection functions, respectively. 
For all data points, the error bars show the 68\% confidence intervals 
assuming the counts in each flux bin are Poisson distributed.}
\end{figure*}
Using this algorithm, the new fit parameters for Equation~(\ref{eq:Tucci1}) are, 
$k_f=$~0.438 $\pm$ 0.0142, 
$n_0=$~46.7 $\pm$ 1.49~Jy$^{-1}{\rm sr}^{-1}$,
$\ell=$~-0.755 $\pm$ 0.137,
and
$S_0=$~2.31 $\pm$ 0.222~Jy.
Note in particular that the addition of our data favors a much smaller value of $k_f$ than the 
value of $k_f=0.5$ presented in T11, which implies a larger number density of faint sources than would 
be inferred by extrapolating the T11 fits to faint fluxes.
The logarithm of the ratio of likelihoods for our best fit parameters and those of T11 is 23.6, 
indicating that our new fit is favored by the data with strong significance.
The selection function minimum flux fits are 
$\smin=$1.48 for the NOAO~DFS catalog and
$\smin=$0.405 for the ATESP catalog.
We use these fits to determine the minimum flux values when creating 
simulated source catalogs in Section~\ref{sec:extrapolation}.

We adopt a simpler maximum-likelihood procedure to fit the steep-spectrum count parameters defined in 
Equation~(\ref{eq:SteepCounts}) for three reasons: 
(1) the steep-spectrum sources presented in the 
left panel of Figure~2 in T11 have much larger uncertainties for constraining the fit function, 
(2) our data turns out to be entirely consistent with the fit function parameters derived from the T11 results, and 
(3) the steep-spectrum number counts are not important for predicting the higher frequency counts at the 
CMB frequencies we are considering in this paper.
The bottom panel in Figure~\ref{fig:flatfit} illustrates how our data are consistent with the T11 result.

\begin{figure}
\begin{center}
\includegraphics[width=0.5\textwidth]{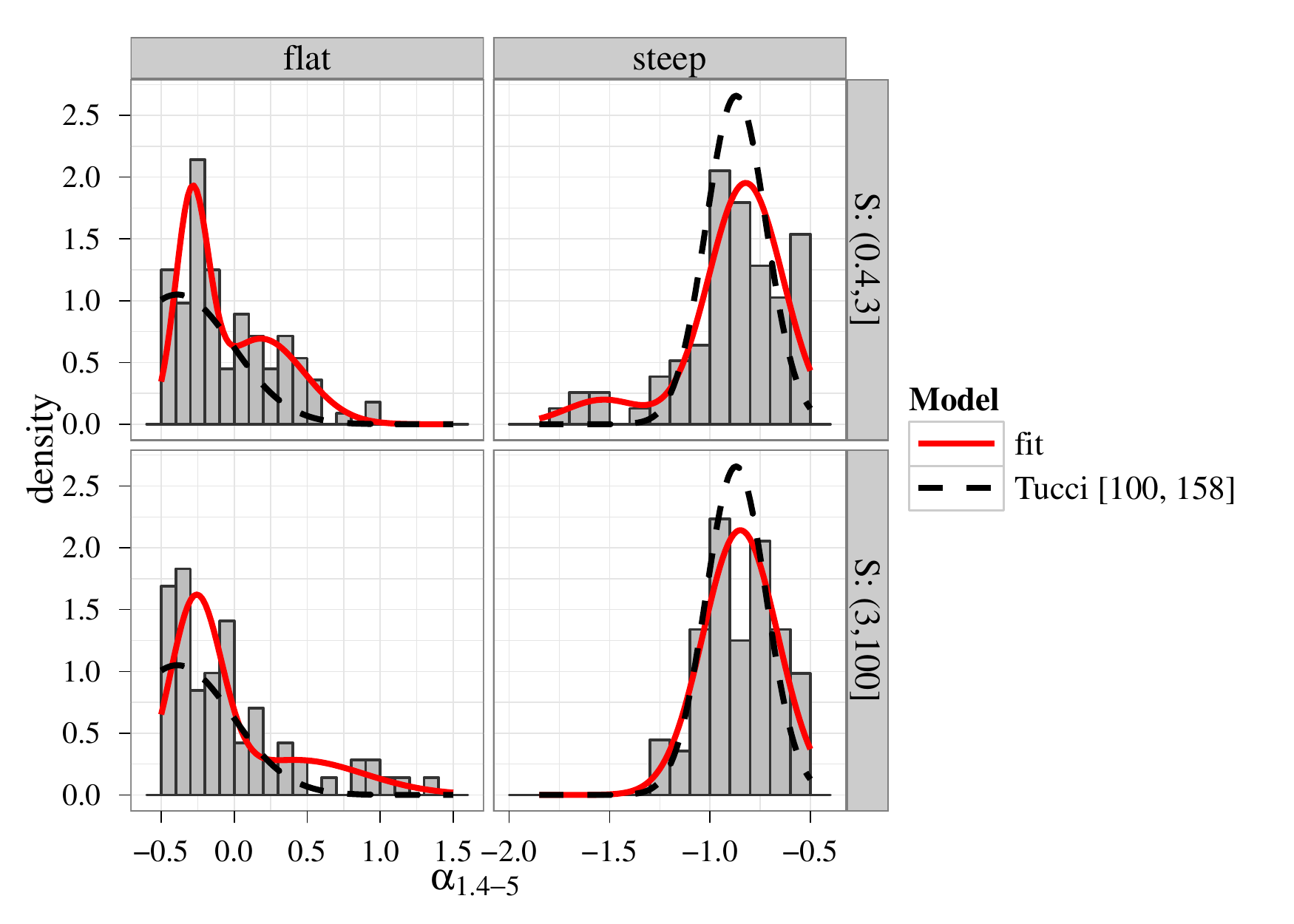}
\caption{\label{fig:indexfit}Distribution of spectral indices for flat$+$inverted (left) and 
steep-spectrum (right) sources in two 5~GHz flux bins (top and bottom rows) 
measured from the combined catalogs of the NOAO~DFS and ATESP fields.
The flux bin ranges in the row side panels are in mJy.
The black dashed lines show the fits from T11 in their $[100, 158]$~mJy flux bin, which is the lowest flux bin they consider.}
\end{center}
\end{figure}

The distributions of spectral indices in 5~GHz flux bins are shown in Figure~\ref{fig:indexfit}.
We make the simplifying assumption that the spectral index is statistically independent of the
5~GHz flux, which is consistent with Figure~\ref{fg:fluxvsindex}.
We again follow~T11 and fit the histograms with separate 
truncated distributions for flat$+$inverted ($\alpha >-0.5$)
and steep ($\alpha \le-0.5$) spectral types (shown by the red solid lines in Figure~\ref{fig:indexfit}).
The dashed black lines in the figure show the fits from T11 for their lowest-flux bin, 
$100 < S /{\rm mJy} < 158$.
We find the distributions are best fit by a mixture of two truncated Gaussian distributions, 
except for the highest-flux steep-source distribution, which is well fitted by a single truncated 
Gaussian.
The parameters of the fits in Figure~\ref{fig:indexfit} are given in Table~\ref{tab:indexfitparams} along with uncertainties on the fit parameters derived 
from the 68\% conditional uncertainty interval on the mean of the largest-index 
Gaussian fit in each panel of Figure~\ref{fig:indexfit}. 
In detail, we calculate uncertainties 
via bootstrap resampling of each catalog, where we also sample the number of catalog entries from a Poisson distribution and then sample with replacement the catalog entries. We then fit double or single Gaussians to each sub-sample shown in Figure~\ref{fig:indexfit}, rank order the fitted Gaussian means, or the larger of the two means when two Gaussians are fit, and select the 68\% confidence intervals from the ordered samples. We focus on the Gaussian means to understand the possible variation (including Poisson uncertainties) in the inverted spectral index distributions.
\begin{deluxetable*}{cclllll}
\tabletypesize{\scriptsize}
  \tablecaption{\scriptsize Fit Parameters for Spectral Index Distributions \label{tab:indexfitparams}}
  \tablewidth{\textwidth}
  \tablehead{ 
  \colhead{Spectral type} & \colhead{Flux Bin} & 
  \colhead{Mean 1} & \colhead{Mean 2} & \colhead{Std. Dev. 1} & \colhead{Std. Dev. 2} & \colhead{Weight} \\
  \colhead{} & \colhead{(mJy)} & \colhead{} & \colhead{} & \colhead{} & \colhead{} & \colhead{}
  }
  \startdata
  flat+inverted & (0.4,3] & $-0.29^{-0.312}_{-0.314}$ & $0.183^{0.257}_{0.149}$ & $0.112^{0.105}_{0.116}$ & $0.295^{0.25}_{0.313}$ & $0.487^{0.556}_{0.479}$ \\ 
  flat+inverted & (3,100] & $-0.262^{-0.234}_{-0.411}$ & $0.428^{0.751}_{0.133}$ & $0.176^{0.207}_{0.0597}$ & $0.46^{0.355}_{0.462}$ & $0.674^{0.806}_{0.41}$ \\ 
  steep & (0.4,3] & $-1.53^{-1.4}_{-1.59}$ & $-0.822^{-0.795}_{-0.844}$ & $0.185^{0.178}_{0.174}$ & $\cdots$ & $0.092^{0.13}_{0.0983}$ \\ 
  steep & (3,100] & $-0.849^{-0.835}_{-0.868}$ & $\cdots$ & $0.186^{0.191}_{0.185}$ & $\cdots$ & $\cdots$\\
  \enddata
\end{deluxetable*}
It is notable for extrapolating fluxes that our best fit distributions for the flat+inverted spectrum 
sources have larger tails with positive spectral indices than the T11 fits in their lowest flux bin.

Note that while~T11 consider how source variability can change the 
inferred spectral indices, their modeling shows that variability has only a small effect on 
the spectral index histograms.  We therefore ignore the effect of variability here.

\subsection{Flux Extrapolation}
\label{sec:extrapolation}
We will primarily use the T11 SED models for extrapolating the 5GHz fluxes to higher frequencies (which 
we will refer to as the ``Tucci SED'' model).  
T11 considered both many different data sets with spectral index information as well as 
physical models for the frequency dependence of the synchrotron emission from the flat-spectrum 
sources to construct statistical SED models that are consistent with observed 
differential number counts at frequencies from 5 to several hundred GHz.  We will show in 
this section that our modified fits to the 5GHz flat+inverted differential counts and the 
spectral index distributions for fluxes less than 100~mJy further improve the fits of the 
extrapolated differential number counts at low fluxes with external data sets.
 
The Tucci SED model assumes that steep-spectrum sources follow a PL SED ($S\propto \nu^\alpha$)
with spectral indices $\alpha_{1.4-5}-\Delta\alpha$ and $\Delta\alpha$ Gaussian distributed to model 
observed steepening at higher frequencies (see Equation~(9) in T11).  
In the description of the steep-spectrum SED,
T11 state that, ``A small percentage of flattening or upturning sources is also included.'' 
We find that we can reproduce their plots of extrapolated steep-spectrum source counts by 
assigning 2\% of our steep-spectrum sources to have spectral indices drawn from a Gaussian 
distribution with mean $-0.3$ and standard deviation 0.2.

We adopt the ``C2Ex'' model from T11 for flat-spectrum sources ($-0.5 <\alpha_{1.4-5} < 0.3$)
that determines the distribution 
of break frequencies according to a physical model of the size of the optically thick synchrotron 
emitting region in FSRQ and BL Lac sources.  The model requires knowledge of the redshift distributions 
of the sources of each type as well as the distributions of Doppler factors that adjust the predicted 
flux from a homogeneous spherically symmetric model to model an asymmetric jet.
Our sample of faint flat-spectrum sources almost certainly has a different redshift distribution 
than that assumed by T11 for their high-frequency predictions.  This is because for a homogeneous 
source sample, fainter fluxes imply higher redshifts, while at fixed redshift fainter fluxes imply sources 
at different evolutionary stage that will again have different redshift distributions.  However, in 
the absence of further information about our catalogs, we simply use the same distributions for the 
flat-spectrum break frequencies as T11 (shown in Figure 7 of T11).  This is probably not 
a terrible approximation because the distributions of the break frequencies span 
several orders of magnitude with the mean following a simple scaling with 5~GHz flux and the 
width determined largely by the assumed size of the emitting region.  So modifications in the 
assumed redshift distribution would yield only minor corrections in the final predicted number counts.

For inverted spectrum sources ($\alpha_{1.4-5} > 0.3$) T11 use the model,
\begin{equation}\label{eq:TucciInvSED}
  S(\nu, \sedparams) = S_0 \left(\frac{\nu}{\nu_0}\right)^{\alpha}
  \left(1 - e^{-\left(\frac{\nu}{\nu_0}\right)^{\lsed-\alpha}}\right),
\end{equation}
where 
\begin{equation}\label{eq:elldist}
  \lsed \sim -A\left(\lsed -\lsed_0\right) \exp\left[ 
  -\frac{1}{2}\frac{\left(\lsed-\lsed_0\right)^2}{\sigma_{\lsed}^2}\right],
\end{equation}
with $\lsed_0=-0.1$ and $\sigma_{\lsed}=0.53$~(see Figure~9 in T11). 
Note that $\lsed$ is therefore defined with support $[-\infty, \lsed_0)$ and
$\lsed_0$ must be less than 0.  The parameter $\nu_0$ in Equation~(\ref{eq:TucciInvSED}) 
is determined from the distribution of peak frequencies in Table~3 of T11.
Again we adopt the distribution for $\lsed$ from T11 without modification.

The T11 SED model as applied in this paper has two main drawbacks. First, the model requires 
specification of both the distributions of ``break'' frequencies where the SEDs of inverted and 
flat-spectrum sources turn over as the radio emission probes different physical regions of the source and
of the distributions of the spectral indices after the break.  
Second, and somewhat related, the T11 SED model has a large number of parameters with unspecified 
uncertainties so that it is difficult to quantify our confidence in the extrapolated 
number counts predicted with this model.
For comparison then, we also consider the physically unrealistic but simple PL SED model,
\begin{equation}\label{eq:sedpl}
  \bar{S}(\nu; S_0, \alpha) = S_0 \left(\frac{\nu}{\nu_0}\right)^{\alpha},
\end{equation}
with $\nu_0 = 5$~GHz.  The PL model will overpredict the number counts at frequencies larger 
than the typical break frequencies for a given source type. But, the PL model has the advantage that we can 
easily propagate our uncertainty in the SED model parameters, $S_0$ and $\alpha$, into the extrapolated 
number count predictions and therefore gain some understanding about the relative uncertainties 
in the number counts from the errors in the data versus the SED model. 

We propagate the uncertainty in the PL SED model by computing the marginal posterior probability distribution 
for the extrapolated flux of each source marginalizing over the SED parameters, $S_0$ and $\alpha$,
and incorporating observational uncertainties in the flux likelihood model. 
We describe the details of this extrapolation method in Appendix~\ref{sec:errorpropagation}.
The marginal posterior probability for the extrapolated flux allows for full uncertainty propagation 
and therefore robust foreground removal (conditioned on the choice of SED model).
Such a procedure becomes less useful when  
the available source catalog is an incomplete sample of the population of sources contributing to 
the foregrounds and when there are many SED parameters to marginalize over leading to large marginal uncertainties.
Our catalogs unfortunately meet the former condition because of the relatively small fraction of the 
sky covered (leading to a dearth of high-flux sources observed) and the Tucci SED model 
likely meets the latter condition.

Putting aside the incompleteness of our catalogs for fluxes $\gtrsim 100$~mJy, we predict the 
high-frequency number counts with the PL SED model using the complete error propagation methods of 
Appendix~\ref{sec:errorpropagation}. Assuming a log-Normal likelihood 
for the observed fluxes, the marginal posteriors for the extrapolated fluxes can be computed analytically
(and therefore quickly). The resulting extrapolated flux confidence intervals can be used as indicators
 of the extrapolation error when the SED model is well-constrained.

However, with the many parameters in the Tucci SED model that, upon marginalization, further contribute 
to the extrapolated flux uncertainties, we 
instead use the T11 flux extrapolation method to study the impact of our new data on the 
``central'' extrapolated count predictions.  In T11 the differential number counts at high frequencies 
are predicted by first simulating a source catalog by sampling sources from the fits to the 5~GHz 
differential counts and sampling spectral indices from the fits to the 5~GHz spectral index histograms, and 
then plugging the sampled values into the SED models for each spectral class.  The extrapolated 
simulated fluxes can be histogrammed to estimate the high-frequency number counts in the same way that the 
observed low-frequency number counts were estimated.
Because the predictions from any one simulated catalog can be subject to random fluctuations (mostly at 
high fluxes where the counts are low) we generate predictions for the Tucci SED model by simulating 10 catalogs 
with $10^6$ sources each and then computing the medians of any summary statistics of the extrapolated catalogs.

\section{High-frequency predictions}
\label{sec:results}
We now apply the flux extrapolation methods described in Section~\ref{sec:extrapolation}
to predict the high-frequency differential number counts and Poisson contributions to 
the CMB power spectrum.
We consider two sets of simulated catalogs for generating predictions with the Tucci SED model.  
First we simulate catalogs with exactly the same parameters as in T11 
(labeled ``Tucci et al. 2011'' in the figures) but with a lower flux limit of 
0.4~mJy, which is where our fit to the ATESP selection function (see Equation~(\ref{eq:selectionfunction}))
has a value of $0.5$. To reach this flux limit we unfairly apply the  
spectral index distributions from T11 for the flux bin [100,158]~mJy to all lower fluxes.  
Our second set of catalogs (labeled ``This work'' in the figures) uses the modified 
$dN/dS$ fit with $\kf=$~0.438 for flat$+$inverted sources as found in Figure~\ref{fig:flatfit} and the 
spectral index distributions fit in Figure~\ref{fig:indexfit} to simulate the spectral indices of sources 
with fluxes lower than the lower limit of 100~mJy in T11.
As described in Section~\ref{sec:extrapolation}, we use the T11 models for both our 
simulated catalogs for the flat and inverted spectrum 
break frequencies and high-frequency spectral indices.

\subsection{Extrapolated Number Counts}
\label{sec:numbercounts}
\begin{figure}
\begin{center}
\includegraphics[width=0.5\textwidth]{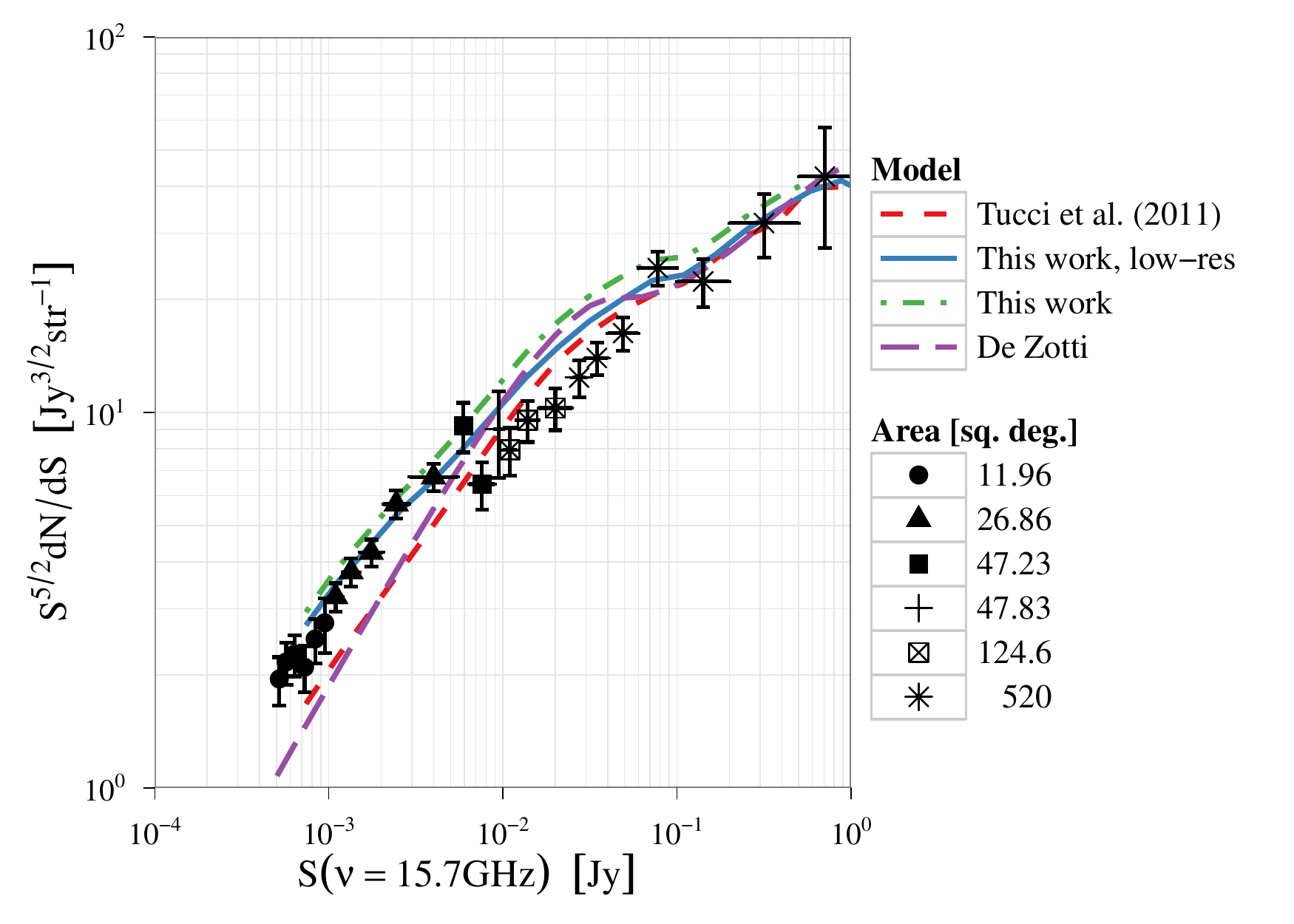}
\caption{\label{fig:Counts15}Comparison of our predicted counts at 15.7~GHz with those measured in the 
9C and 10C surveys by~\citet{2011MNRAS.415.2708A}.  
The dash-dotted green line shows our model prediction while the solid blue line shows the same 
model prediction after merging sources to be consistent with the 30 arcsecond resolution of the 9C and 10C surveys. In both cases we draw sources from our fits
to the 5~GHz counts and 1.4--5~GHz spectral index distributions and then 
extrapolate the fluxes to 15.7~GHz (before merging the sources for the solid 
blue line). The dashed red line instead uses the 5~GHz counts and index 
distribution fits provided in T11 
(unfairly applied at fluxes lower than those considered in T11).
The long-dashed purple line shows the model from~\cite{2005A&A...431..893D}.
The data points are from Table~6 of \citet{2011MNRAS.415.2708A} with Poisson error bars added.}
\end{center}
\end{figure}

To validate the differential counts model based on our catalogs we first compare with the 
measured differential counts at 15.7~GHz in the 9C and 10C surveys~\citep{2011MNRAS.415.2708A} in 
Figure~\ref{fig:Counts15}.
The combined 10C survey covers 12 deg$^2$ to a flux limit of 0.5~mJy while the 9C survey data 
cover several hundred square degrees to a limit of 10~mJy.  While this provides excellent statistics 
for the estimates of the differential number counts, the surveys only have a resolution of 30 arcsec 
compared to our catalogs that have a resolution of 5 arcsec.  
It is therefore possible that the number of sources observed with 30 arcsec resolution is artificially 
reduced relative to our model because resolved sources at 5 arcsec are merged together 
when they happen to fall close together in projection on the sky.
Out of the 362 sources in our NOAO~DFS catalog, we find  
29 source pairs that lie within 30 arcsec of each other and 
10 source triplets within a circle of 30 arcsec diameter.
To predict the 15.7~GHz differential counts for 30~arcsec resolution we 
randomly choose sources to merge in our simulated catalogs so that the fractions of merged 
pairs and triplets match those in the NOAO~DFS catalog. Because of low statistics we have made 
no attempt to choose sources to merge based on any other criteria than random selection.
The resulting predicted number counts are shown by the solid blue line labeled ``This work, low-res''
in Figure~\ref{fig:Counts15}.  Our model is a much better match to the 10C survey number counts
than either the T11 model or the \cite{2005A&A...431..893D} model for fluxes less than 8~mJy.
The better fit at faint fluxes is mostly due to the increased 5~GHz counts from the fit we 
found in Figure~\ref{fig:flatfit}.
For the fluxes from 10~mJy to 60~mJy the 9C survey number counts fall well below all of the model 
predictions, while all the models are in good agreement with the observed counts for larger fluxes.

\begin{figure}
\begin{center}
\includegraphics[width=0.5\textwidth]{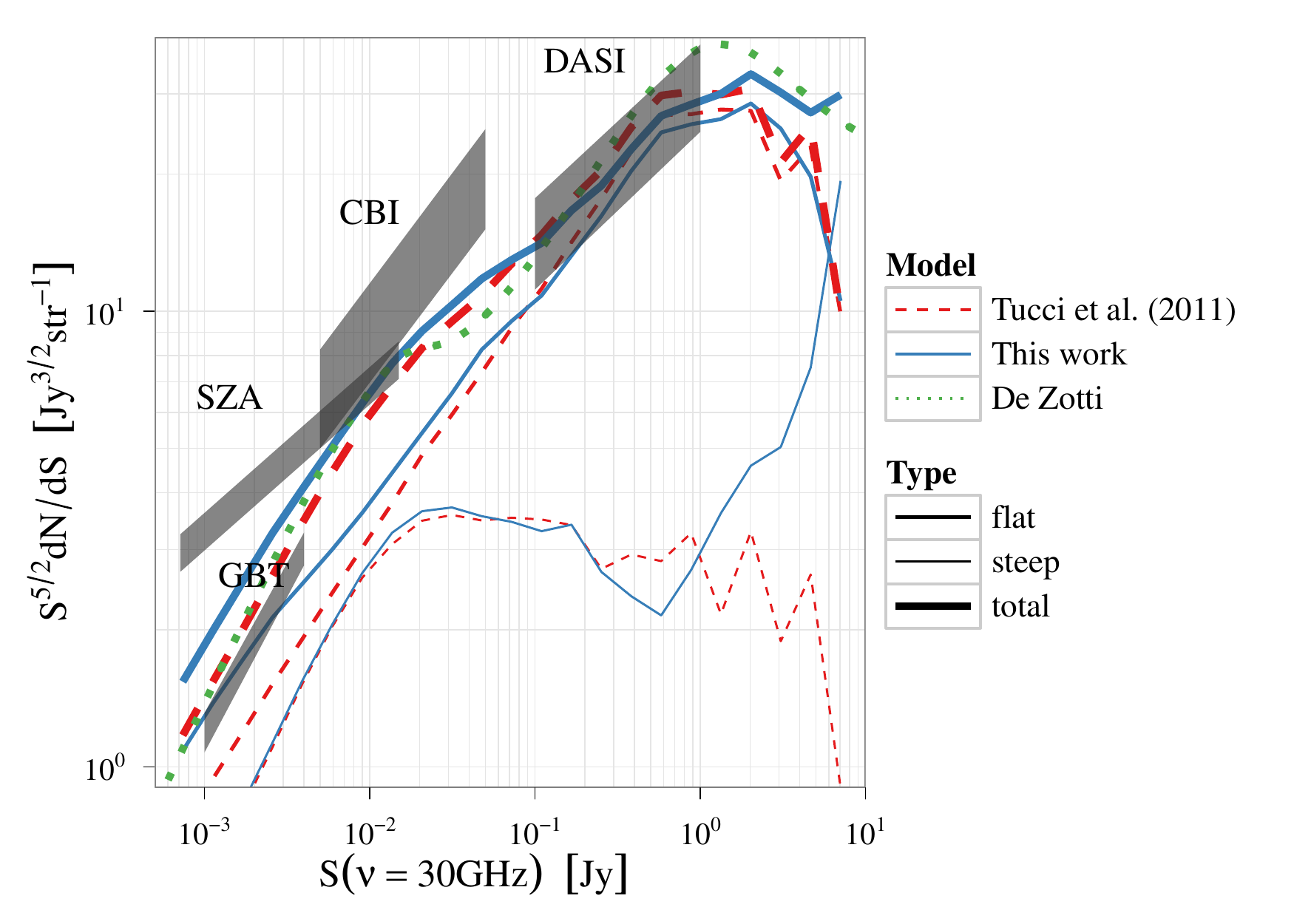}
\caption{\label{fig:counts33}Predicted differential number counts at 30~GHz 
using the T11 SED model to extrapolate the modeled 5~GHz counts. 
The thin lines show the steep source counts (at the bottom of the figure),
the slightly thicker lines show the flat+inverted counts, and the 
thickest lines show the sum of the steep and flat+inverted counts. 
The solid blue lines use our fits to the 5~GHz counts and 1.4--5~GHz spectral index 
distributions while the dashed red lines use those from T11 (as in Figure~\ref{fig:counts15}).
The dotted green line is the model from~\cite{2005A&A...431..893D}.
The shaded bands showing measurements at 33~GHz are copied from Figure 12 in T11.
}
\end{center}
\end{figure}

In Figure~\ref{fig:counts33} we recreate Figure~12 from T11 to compare our predicted differential number counts
at 30~GHz with several other surveys.  
The steep-spectrum sources are a small but non-negligible contribution to the number counts at 30~GHz.  
The new fit to the low-flux slope of the 5~GHz differential counts in Figure~\ref{fig:flatfit} and 
the larger values of the spectral indices shown in Figure~\ref{fig:indexfit}
lead to a significant difference in the total 30~GHz counts for fluxes between 1 and $\sim50$~mJy.  
Our fit leads to greater consistency than the T11 model with the counts measured by  
the Cosmic Background Imager (CBI)~\citep{2003ApJ...591..540M} and the 
Sunyaev--Zel'dovich Array (SZA)~\citep{2010ApJ...716..521M},
but is in greater disagreement with the Green Bank Telescope (GBT) 
survey by~\citet{2009ApJ...704.1433M} (all of which were previously presented and compared in T11).
The disagreement between our predictions and the GBT survey measurements might be explained in part 
by the lower resolution of 24 arcsec for the GBT survey as well as the targeted source selection 
from the NVSS.
For all fluxes less than $\sim 0.1$~Jy our predicted counts are now larger than the \citet{2005A&A...431..893D}
model (shown by the dotted green line).
Also note that our predicted number counts are significantly larger than those of~\citet{2010MNRAS.407..247C} 
(their Figure~5) from a simulation based on the NVSS source catalog.

\begin{figure}
\begin{center}
\includegraphics[width=0.5\textwidth]{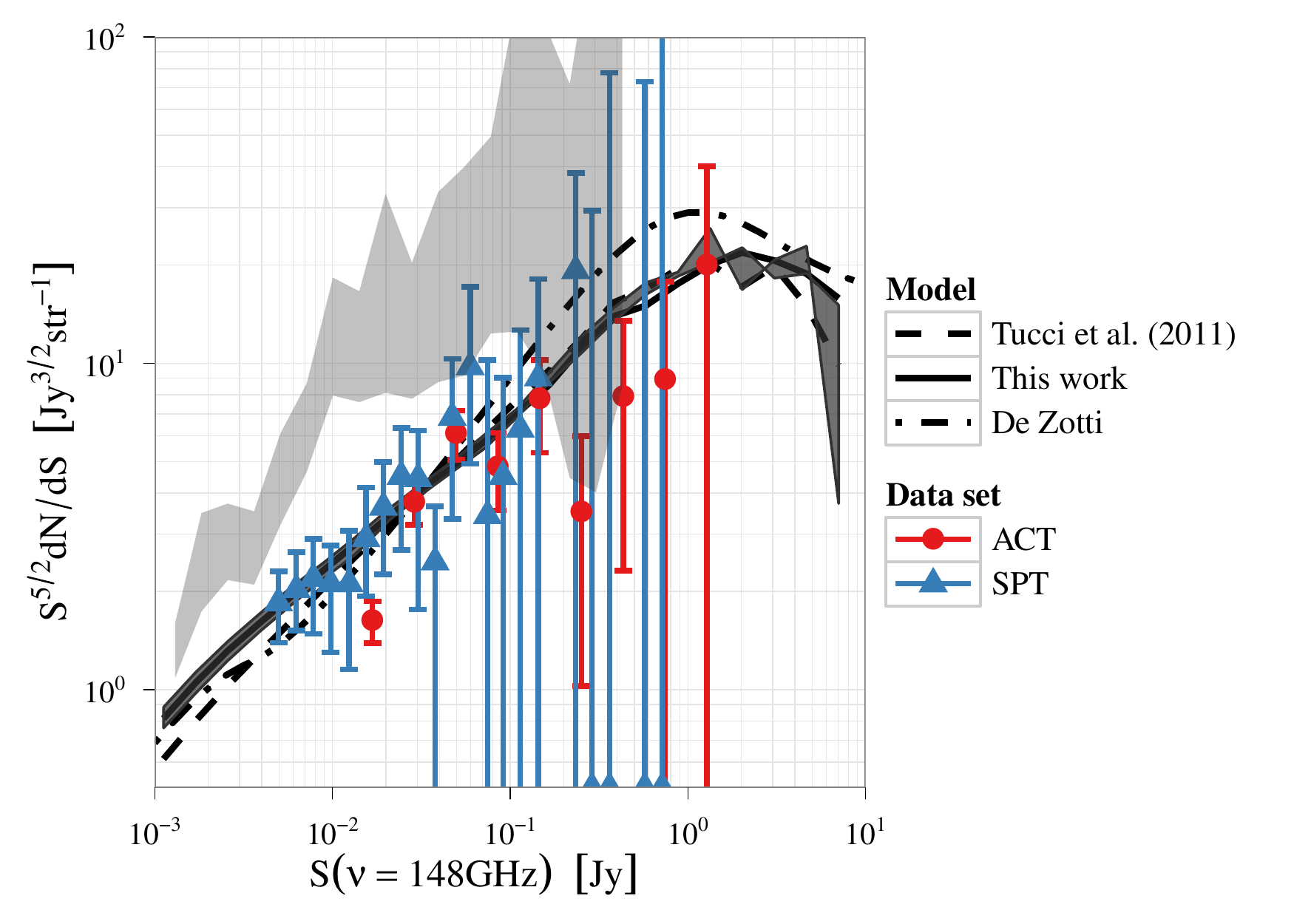}
\caption{\label{fig:counts148}Predicted differential number counts at 148~GHz with the same models as in 
Figures~\ref{fig:counts15} and \ref{fig:counts33}. 
The dark gray shaded band denotes the range of extrapolated number count 
predictions when the faint slope, $k_f$, for the flat+inverted 5~GHz number counts 
is varied over its 68\% confidence interval.
The data points 
are taken from \citet{2011ApJ...731..100M} for ACT and \citet{2010ApJ...719..763V} for SPT.
The light gray shaded band shows the 95\% confidence intervals for the number counts when each source flux in the 
NOAO~DFS and ATESP catalogs is extrapolated using a power-law SED as described in Appendix~\ref{sec:errorpropagation}. The power-law SED band is intended to 
show the range of extrapolated count uncertainties when marginalizing 
over SED parameters given only the NOAO~DFS and ATESP catalogs.}
\end{center}
\end{figure}

Finally, ACT~\citep{2011ApJ...731..100M} 
and SPT~\citep{2010ApJ...719..763V} have recently 
presented catalogs of resolved point sources at 148~GHz, 
which we compare with our model in Figure~\ref{fig:counts148}. 
Again, our model 
predicts larger number counts than the T11 and De~Zotti models 
for fluxes less than $\sim50$~mJy.  
The uncertainty 
in the predicted counts from the 68\% uncertainty range in $k_f$ is shown by the 
gray shaded band in Figure~\ref{fig:counts148} and is much less than the difference 
between models for $S\lesssim50$~mJy.
All the models are consistent with the SPT measurements
but the increased counts for faint fluxes predicted by our model is in tension with the 
lowest-flux ACT measurement. 
The model predictions at 148~GHz depend sensitively on the SED model 
parameters, in particular the flat-spectrum source break frequency distribution 
described in T11 and Section~\ref{sec:extrapolation}. So it is possible that the tension 
between our model prediction and the ACT measurement indicates that the break frequencies are 
really smaller than we have modeled for the 
flat-spectrum sources in our catalog with 5~GHz fluxes less than 10~mJy.
However, the break frequencies are degenerate with the spectral indices after the break in the SED 
when comparing with high-frequency number counts so we cannot make unambiguous conclusions about the 
SED model from the comparison in Figure~\ref{fig:counts148}.

We have also plotted the 95\% confidence intervals for the 
extrapolated number counts as a gray shaded band in Figure~\ref{fig:counts148} 
assuming a PL SED model and using the error propagation methods described in 
Appendix~\ref{sec:errorpropagation}. The confidence intervals 
for the PL SED model include both uncertainties in the SED parameters and the 
Poisson uncertainties from the limited size of the NOAO~DFS and ATESP catalogs (the PL SED 
predictions use only the NOAO~DFS and ATESP catalogs as input). When extrapolating 
source fluxes over such a broad frequency range the PL SED is of course a poor model, but 
we include it in Figure~\ref{fig:counts148} to demonstrate the expected modeling uncertainties 
for the flux extrapolation.  A similar error propagation for all the SED parameters in the Tucci 
model could yield even larger uncertainties. But the Tucci SED model is calibrated against many 
other datasets so the priors on the SED parameters might limit the increase in the uncertainties 
relative to the PL SED model when all SED parameters are marginalized.  We have left this 
investigation for future work.

\subsection{CMB Power Spectrum}
\label{sec:cmb}
As shown in, e.g.,~\citet{1996MNRAS.281.1297T,1999AA...346....1S} the 
Poisson contribution to the CMB power spectrum from unresolved point sources is
\begin{equation}\label{eq:clpoisson}
  C_{\ell}(\nu) = g^2(\nu)\int_{0}^{\scut} S_{\nu}^2 \frac{dN}{dS_{\nu}}\, dS_{\nu},
\end{equation}  
where $g(\nu)$ converts from intensity at frequency $\nu$ to temperature in $\mu$K, 
and $\scut$ is the minimum flux (in a given channel) at which point sources can be 
resolved and masked out or otherwise removed.
The same radio sources also contribute a clustering foreground to the CMB power spectrum, 
but this is expected to be sub-dominant to the Poisson power for the source intensities 
we are considering here~\citep{1999AA...346....1S}.

In Table~\ref{tab:clplanck} we evaluate Equation~(\ref{eq:clpoisson}) for the lower frequency 
Planck channels using the $\scut$ values from the Early Release Compact Source catalog (ERCSC) 
given in Table~3 of 
\citet{PlanckERCSC}. We have omitted predictions for the Poisson power for the three highest 
frequency Planck channels both because the unresolved radio sources become a sub-dominant 
foreground at those frequencies and because the flux extrapolation becomes increasingly 
unreliable. 
  
  
  
\begin{deluxetable}{cccc}
\tabletypesize{\scriptsize}
  \tablecaption{\scriptsize Predicted Poisson Power for Planck Lower-frequency Channels \label{tab:clplanck}}
  \tablewidth{0.5\textwidth}
  \tablehead{
  \colhead{Freq.} & \colhead{$\scut$} & 
  \colhead{$C_{\ell}$ Tucci} & \colhead{$C_{\ell}$ This Work} \\
  \colhead{(GHz)} & \colhead{(Jy)} &
  \colhead{($\mu {\rm K}^2$)} & \colhead{($\mu {\rm K}^2$)}
  }
  \startdata
  30 & 0.48 & 2.8E-02 & 2.7E-02 \\ 
  44 & 0.58 & 6.6E-03 & 6.2E-03 \\ 
  70 & 0.48 & 8.2E-04 & 7.7E-04 \\ 
  100 & 0.34 & 1.6E-04 & 1.5E-04 \\ 
  143 & 0.21 & 3.3E-05 & 3.2E-05 \\ 
  217 & 0.18 & 1.5E-05 & 1.5E-05 \\ 
  \enddata
\end{deluxetable}
We show similar predictions for the Poisson power for ACT and SPT in Table~\ref{tab:clspt} 
assuming $\scut=0.005$~Jy for SPT~\citep{2010ApJ...719..763V} and 
$\scut=0.02$~Jy for ACT~\citep{2011ApJ...731..100M}.
\begin{deluxetable}{cccc}
  \tabletypesize{\scriptsize}
  \tablecaption{\scriptsize Predicted Poisson Power for ACT and SPT \label{tab:clspt}}
  \tablewidth{0.5\textwidth}
  \tablehead{
  \colhead{Freq.} & \colhead{$\scut$} & 
  \colhead{$C_{\ell}$ Tucci} & \colhead{$C_{\ell}$ This Work} \\
  \colhead{(GHz)} & \colhead{(Jy)} &
  \colhead{($\mu {\rm K}^2$)} & \colhead{($\mu {\rm K}^2$)}
  }
  \startdata
  95 & 0.005 & 2.5E-06 & 3.1E-06 \\ 
  148 & 0.005 & 6.1E-07 & 7.6E-07 \\ 
  148 & 0.02 & 2.7E-06 & 3.1E-06 \\ 
  217 & 0.005 & 3.2E-07 & 4.0E-07 \\ 
  217 & 0.02 & 1.5E-06 & 1.7E-06 \\ 
  277 & 0.02 & 1.6E-06 & 1.8E-06 \\ 
  \enddata
\end{deluxetable}

\begin{figure*}
\centerline{
  \includegraphics[width=0.65\textwidth]{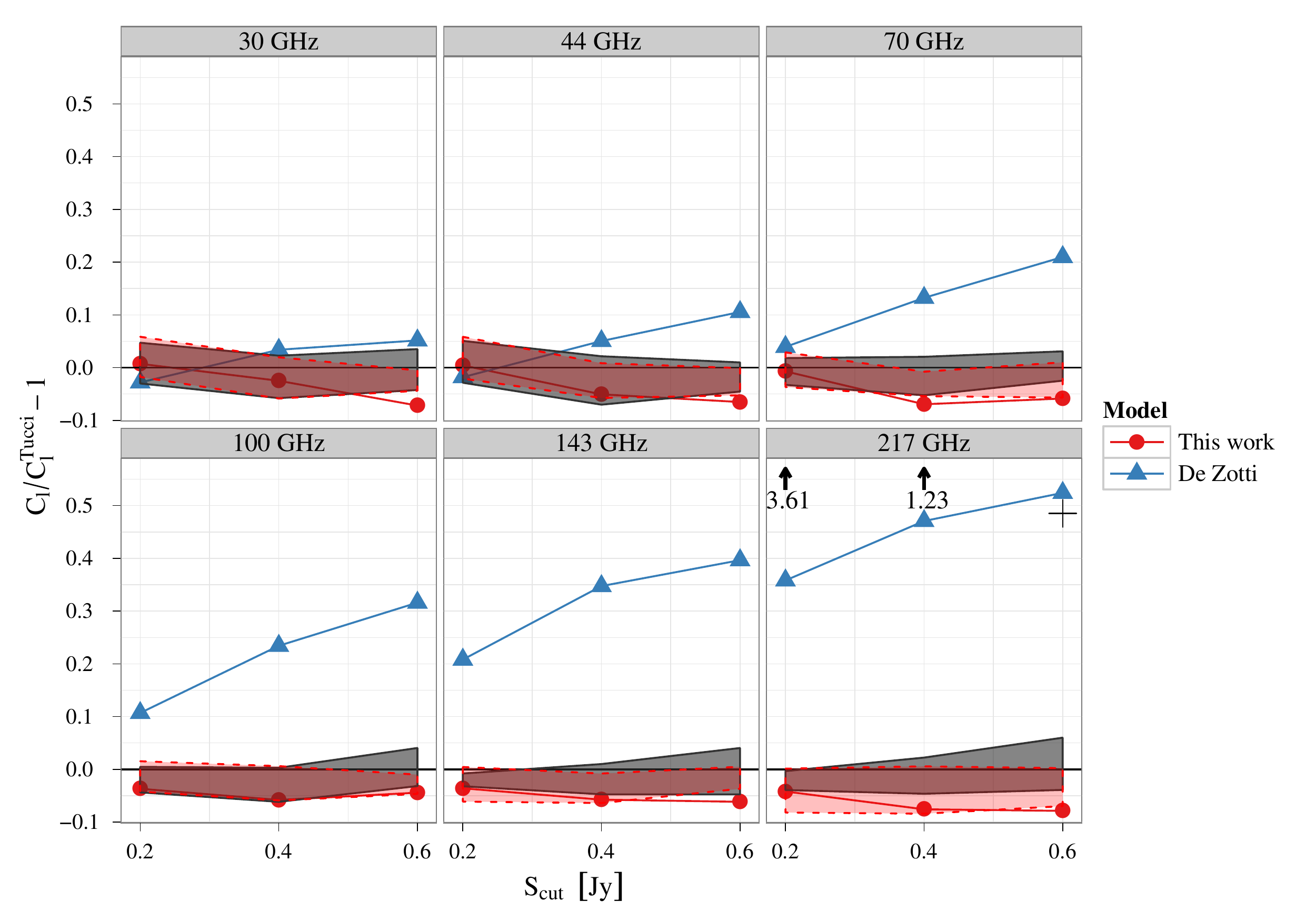}
}
\centerline{
  \includegraphics[width=0.65\textwidth]{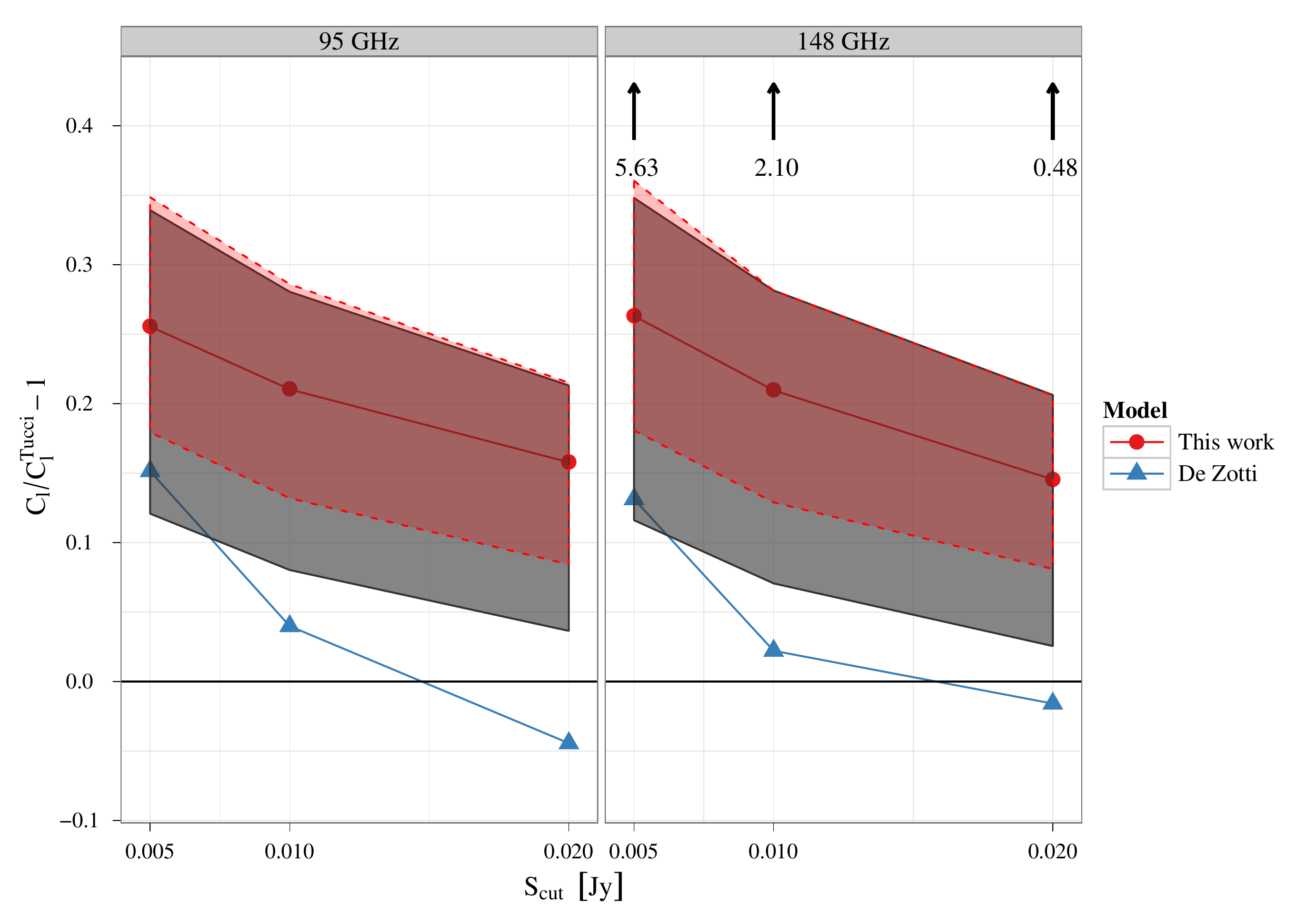}
}
\caption{\label{fig:Clfraction} Predicted CMB Poisson power from unresolved radio point sources as a function of the minimum 
flux, $\scut$, at which point sources can be resolved and excised. The model from this paper (``This work'') and the model 
from \cite{2005A&A...431..893D} are shown by the red circles and blue triangles, respectively, normalized by the model from 
T11. The red shaded bands with the dashed outlines show the range of the predicted Poisson power when the flat+inverted 5~GHz 
differential number count slope $k_f$ (see Equation~(\ref{eq:Tucci1})) is varied over its 68\% confidence interval (when fit 
to our data and that in T11). The gray shaded bands show the range of predicted Poisson power including both the aforementioned 
variation in $k_f$ as well as the variation of the 1.4--5~GHz spectral indices over the 68\% conditional confidence intervals in 
the mean of the inverted spectral index distributions (see Figure~\ref{fig:indexfit} and Table~\ref{tab:indexfitparams}).
At frequencies where the Poisson power from infrared sources is comparable or larger than that from radio sources we have included 
either a black cross or black arrows (when the T11 normalized values do not fit within the plot range). In both the 
top and bottom panels, the millimeter source Poisson power dominates the radio source power for higher frequencies than those shown 
but is highly sub-dominant for all but the highest shown frequency panels.}
\end{figure*}

We further compare the different model predictions for the Poisson power in Planck, ACT and SPT channels 
in Figure~\ref{fig:Clfraction}. Each panel shows the predicted Poisson power 
for our model and the De~Zotti model, both normalized to the prediction from the T11 model as a 
functions of $\scut$.  The top six panels show the predictions for the six lowest-frequency Planck channels
while the bottom two panels show predictions for the lower ACT and SPT frequencies and $\scut$ ranges.
We have plotted only those frequency channels that have predicted radio source Poisson power much larger or comparable to the infrared source Poisson power as predicted by the model from Equation~(4) of~\cite{2011arXiv1102.5195M} (using their amplitude normalized to Planck and SPT). We plot the predicted infrared source Poisson power with black crosses or arrows (when outside the scale of the panel) only where the prediction is comparable to or larger than the radio source Poisson power (i.e. the highest frequency panel for each $\scut$ range).
The circular points and red lines in Figure~\ref{fig:Clfraction} show the predictions using the mean of the marginal posterior for $k_f$ from Equation~(\ref{eq:Tucci1}) while the blue triangles and lines show the model from~\cite{2005A&A...431..893D}.
The red shaded bands show the range of predictions for $k_f$ values spanning the 68\% confidence intervals 
of the marginal posterior given the 5~GHz number count observations. The gray shaded bands are similar, but also 
include the variation of the 1.4--5~GHz spectral indices of the flat+inverted sources over their 68\% confidence intervals
as reported in Table~\ref{tab:indexfitparams}.
For Planck, our model predicts Poisson power that matches the prediction from the T11 model to within 10\% for 
all $\scut$ values. This is because our data modify the T11 model at 5~GHz only for source fluxes 
$\lesssim100$~mJy while the Planck Poisson power is mostly determined by sources with fluxes just below $\scut>100$~mJy.
The De~Zotti model predicts higher Poisson power than the T11 model and our model because of an 
excess of sources with fluxes just below $\scut$. This is the reason for the different 
dependence on $\scut$ of the two models in the 70 and 100~GHz panels of Figure~\ref{fig:Clfraction}.
At the ACT and SPT frequencies in the bottom 2 panels of Figure~\ref{fig:Clfraction}, our model predictions
are consistently above both the De~Zotti and T11 models, as also indicated in Figure~\ref{fig:counts148}.
The excess faint sources in our catalogs predict increased Poisson power in ACT and SPT of 5\%-30\% 
relative to the T11 model, ignoring any uncertainties in the Tucci SED model parameters.

\begin{figure*}
\begin{center}
\includegraphics[width=0.7\textwidth]{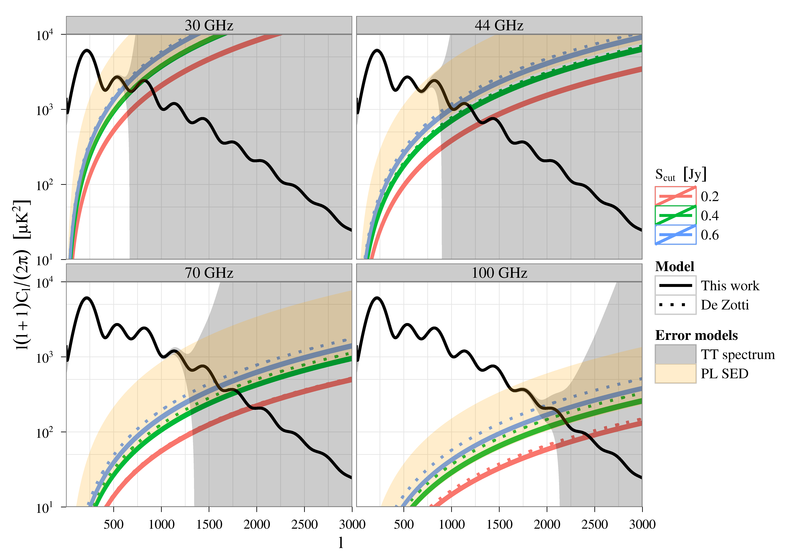}
\caption{\label{fig:Cl}CMB power spectra in the four lowest-frequency Planck channels. 
The Poisson contribution to the power from unresolved point sources is shown for three 
different values of $\scut$ using both the updated 5~GHz number count model and 
spectral index distributions (``This work'') and the model from \citet{2005A&A...431..893D}. The line widths for our model (``This work'') denote 
the prediction range when $k_f$ is varied over its 68\% confidence interval. 
The T11 model predictions are always within the line widths of our model.
The black line and gray shading show the TT CMB power spectrum 
and expected errors from pixel noise and beam smearing.  The orange shaded band shows the 
95\% credible intervals for the Poisson contribution from the sources in the NOAO~DFS catalog and 
assuming a power-law (``PL'') SED model with $\scut=0.4$~Jy. }
\end{center}
\end{figure*}

In Figure~\ref{fig:Cl}, we plot the different Poisson power model predictions 
in the four lowest-frequency Planck channels 
compared with the TT
CMB power spectrum \citep[predicted with Emu~CMB][]{2011ApJ...728..137S} 
and the expected errors from the average noise per pixel and beam smearing.
Figure~\ref{fig:Cl} makes it clear that the Poisson contribution to the total 
observed power is comparable to or sub-dominant to the noise in each channel.
While the resolution is better in higher-frequency Planck channels, the Poisson 
power from millimeter sources will dominate over the radio sources we are modeling.
We also plot in Figure~\ref{fig:Cl} the 95\% confidence intervals on the predicted Poisson power 
assuming a PL SED derived using the methods in Appendix~\ref{sec:errorpropagation}. 
The sizes of the uncertainties for the PL SED are comparable to the variation in the Poisson power 
for the range of $\scut$ values in the ERCSC and are much larger than the variation between 
the Tucci SED models for a fixed $\scut$ value.  

We showed in the upper panels of Figure~\ref{fig:Clfraction} that our 5~GHz number count 
measurements predict no more than 10\% changes in the predicted radio source Poisson power in any 
Planck channel relative to earlier models. We now consider whether our data 
yield 5~GHz count measurements that provide a sufficient model to remove the bias in constraints 
on the scalar spectral index, $n_s$, from radio foregrounds in the Planck 
temperature power spectrum.
We adopt the Fisher matrix formalism as described in~\citet{2005APh....23..369H} to predict the 
bias on $n_s$ assuming our model with the 5~GHz number count slope $k_f=0.452$ (our $+1\sigma$ fit value) is the ``truth,'' but 
that our model with $k_f=0.423$ (our $-1\sigma$ fit value) is used to subtract the Poisson power when analyzing the data.  
For computing the Fisher matrix we use the power spectrum noise model including the Poisson 
contribution from \citet{1998PhRvD..57.5273W} as well as the optimal linear estimator therein 
for the CMB power spectrum given the five lowest-frequency Planck channels.  
The inferred biases on $n_s$ assuming the same $\scut$ value for each channel are shown 
in Table~\ref{tab:nsbias}.  The last column of the table also shows the 
biases normalized by the marginal Fisher matrix 1-$\sigma$ uncertainties on $n_s$.
\begin{deluxetable}{ccc}
\tabletypesize{\scriptsize}
  \tablecaption{\scriptsize Scalar Spectral Index Bias from Wrong Point-source Model with 
  Four Low-frequency Planck Channels.\label{tab:nsbias}}
  \tablewidth{0.5\textwidth}
  \tablehead{
  \colhead{$\scut$} & \colhead{$n_s$ Bias} & \colhead{$n_s$ Bias$/$$\sigma(n_s)$} \\
  \colhead{(Jy)} & \colhead{} & \colhead{}
  }
  \startdata
  0.6 & 0.0080 & 0.55 \\ 
  0.4 & 0.0071 & 0.49 \\ 
  0.2 & 0.0037 & 0.25 \\ 
  \enddata
\end{deluxetable}
We marginalized over five other cosmological parameters~\citep[as in][]{2011ApJ...728..137S}.
For all $\scut$ values, the biases on $n_s$ are smaller than the 1$\sigma$
marginal Fisher errors.  So, by this measure, the remaining uncertainties in the 5~GHz counts 
fit from our catalogs are not significant for Planck parameter estimation (neglecting uncertainties in the SED models).

\section{Discussion}
\label{sec:conclusions}
We have shown that a previously unconsidered population of faint radio sources 
at 1.4 and 5~GHz with flat and inverted spectral indices could significantly 
contribute to the unresolved point source contamination in the power spectrum
measurements in ongoing high resolution CMB experiments. However, the measured 
differential number counts of faint radio sources are not large enough to significantly 
modify previous predictions for the Poisson power in the Planck temperature power 
spectrum.
Our source catalog is unique in the combination of flux sensitivity to 1.5~mJy, 
resolution of 5~arcsec, area covered of 7~deg$^2$ and matching of sources in the 
1.4 and 5 GHz frequency bands. 

We found that existing fits to the differential number counts of 5~GHz flat and inverted spectrum 
sources must have a shallower slope for fluxes less $\sim100$~mJy in order to be consistent with our
catalog.  We also found that the distribution of 1.4--5~GHz spectral indices of sources with 5~GHz 
fluxes less than 100~mJy is skewed toward larger values than the distributions for higher flux sources.
Taken together these changes to the 5~GHz source counts imply increased differential counts when 
the fluxes are extrapolated to higher frequencies and therefore increases in the predicted 
level of Poisson power from unresolved point sources in CMB measurements.

The quantitative predictions of high-frequency number counts depend on the choice of SED model 
for extrapolating the measured 5~GHz fluxes. And the uncertainties in the predicted high-frequency 
counts depend on the prior constraints on SED model parameters given a choice of SED model.
Our main results assume a physically motivated SED model \citep[from][]{2011A&A...533A..57T}
but we also investigated a simple PL SED model to assess the dependence of our
predictions on SED modeling uncertainties.
In general we found that propagating the uncertainties from both the limited size of our catalog 
and two unknown SED model parameters can lead to high frequency prediction uncertainties that are much 
larger than the differences between the predictions of competing SED models.
However, we have argued that systematic offsets in the high-frequency predictions may persist 
even when the prediction uncertainties are large so that the predictions calibrated against our 
catalog will be important considerations for analyzing CMB experiments.  
As one example we find that if the four lowest-frequency Planck channels are optimally combined 
to constrain $n_s$ from the CMB temperature power spectrum, the constraints are biased by less than the size of the 1$\sigma$ uncertainties if the ``wrong'' point source number count model is used 
to subtract the Poisson power. We limited our forecasts in this example to the five lowest 
frequency Planck channels to avoid modeling other more dominant foregrounds at higher frequencies.  
So, our forecasts are simply an illustration of the significance of the difference between point source 
models rather than predictions of the optimal Planck constraints when all channels are considered, but we argue that our catalog shows that radio source models at 
5~GHz are sufficient for removing Planck Poisson power as a systematic.

We outlined methods in Appendix~\ref{sec:errorpropagation} for propagating both
measurement and SED model uncertainties when extrapolating fluxes on a source-by-source
basis. A rigorous approach to predicting the Poisson power from unresolved point
sources in CMB observations could apply the methods of
Appendix~\ref{sec:errorpropagation} to a combined catalog of all available radio
survey data at relevant frequencies. We expect that significant constraints on the
uncertain SED model parameters could be imposed by such a joint analysis of available
data, thereby removing a key uncertainty from the present work while also providing 
useful physical parameterizations of the radio Poisson 
foreground~\citep[as used in][]{2011arXiv1112.3260P}. However, we leave this for future investigations.

\section*{Acknowledgments}
We thank Lloyd Knox, Marius Millea, and Mark Ammons for helpful comments on an early version of this draft and an 
anonymous referee for significant improvements.
This work performed in part under the auspices of the U.S. Department of Energy by 
Lawrence Livermore National Laboratory under Contract DE-AC52-07NA27344.
The National Radio Astronomy Observatory is a facility of the National Science Foundation operated under cooperative agreement by Associated Universities, Inc.

{\it Facility:} \facility{VLA}.

\appendix
\section{Error propagation for extrapolated fluxes}
\label{sec:errorpropagation}
For a given SED model $\bar{S}(\nu, \sedparams)$ with parameters $\sedparams$, 
we assume a log-normal likelihood for the observed fluxes of the $i$th 
object in the catalog.
We choose a log-normal likelihood because it has strictly positive support 
and becomes nearly symmetric about the mean when the flux errors are small 
or the flux is large.

To propagate our uncertainty in the measured fluxes and SED parameters to 
the extrapolated fluxes, we compute the marginal posterior distribution 
for the extrapolated flux,
\begin{equation}\label{eq:marggeneral}
  P(S_{\nu} | \sobs_{i}) = \int d\sedparams\, P(S_{\nu} | \sedparams)\,
  P(\sedparams | \sobs_{i}).
\end{equation}
The first term gives a delta function, $P(S_{\nu}|\sedparams) = 
\delta_{D}(S_{\nu} - \bar{S}(\nu, \sedparams))$, which fixes one of the 
SED parameters as a function of the remaining SED parameters and the given 
value of $S_{\nu}$.  For a PL SED, the delta 
function effectively sets 
$\alpha = \ln\left(\frac{S_{\nu}}{S_0}\right) / \ln\left(\frac{\nu}{\nu_0}\right)$ 
(or $S_0=S_0(S_{\nu},\alpha)$).  
The marginalization in Equation~(\ref{eq:marggeneral}) then becomes
\begin{equation}
  P(S_{\nu} | \sobs_{i}) \propto \frac{1}{S_{\nu}\ln(\nu/\nu_0)}
  \int dS_0\, P(\sobs_i | S_0, \alpha(S_0, S_{\nu})) P(S_0),
\end{equation}
where the coefficient in front of the integral comes from the transformation of 
variables in the delta function. We assume a conjugate log-normal prior for $S_0$ with 
mean $\mu_{S_0}$ and standard deviation $\sigma_{S_0}$.

Performing the integration over $S_0$, the final expression for the 
marginal posterior for the extrapolated flux of an individual source for a 
PL~SED is
\begin{equation}
  P(S_{\nu} | \sobs_i) = \frac{1}{A_{S_{\nu}}} S_{\nu}^{\alpha_{S_{\nu}}}
  \exp\left[- \frac{1}{2} 
  \frac{\left(\ln(S_{\nu}) - \mu_{S_{\nu}}\right)^2}
  {\sigma_{S_{\nu}}^2}\right],
\end{equation}
where, if we assume $\nu_0=1.4$~GHz,
\begin{equation}
  x_5 \equiv \frac{\ln(5 / \nu_0)}{\ln(\nu / \nu_0)},
\end{equation}
and with $\sigma_{1,5} \equiv \sigma_{S_{1.4,5}} / S_{1.4, 5}$,
\begin{equation}
  \mu_{S_{\nu}} \equiv x_5 
  \frac{(\sigma_{S_0}^2 + \sigma_1^2)\ln(S_5) + (x_5-1)\sigma_{S_0}^2\ln(S_1)}
  {x_5^2(\sigma_{S_0}^2 + \sigma_1^2)},
\end{equation}
\begin{equation}
  \sigma_{S_{\nu}} \equiv 
  \frac{\sigma_1^2 \sigma_5^2 + \sigma_{S_0}^2 ((x_5-1)^2 \sigma_1^2 + \sigma_5^2)}
  {x_5^2 (\sigma_{S_0}^2 + \sigma_1^2)},
\end{equation}
\begin{equation}
  \alpha_{S_{\nu}} \equiv 
  -1 + \frac{\mu_{S_0} (x_5-1) x_5 \sigma_1^2}
  {\sigma_1^2 \sigma_5^2 + \sigma_{S_0}^2 ((x_5-1)^2\sigma_1^2 + \sigma_5^2)},
\end{equation}
and, 
\begin{equation}
  A_{S_{\nu}} \equiv \sqrt{2\pi \sigma_{S_{\nu}}^2} 
  \exp\left[\frac{1}{2} (1 + \alpha_{S_{\nu}})^2
  2\mu_{S_{\nu}} + (1 + \alpha_{S_{\nu}}) \sigma_{S_{\nu}}^2\right].
\end{equation}

We then estimate the mean differential number counts by summing 
over the (normalized) posteriors in bins in flux,
\begin{align}
  \frac{dN_k(>S)}{dS} \Delta S_k &\approx \sum_{i=1}^{N_{\rm sources}}
  \int_{S_{{\rm min},k}}^{S_{{\rm max},k}}  dS\, P(S_{\nu} | \sobs_{i})
  \notag\\
  &\equiv \sum_{i=1}^{N_{\rm sources}}\, p_{ik},
\end{align}
where $\Delta S_k\equiv S_{{\rm max},k} - S_{{\rm min},k}$ and $k=1,\dots,N_{\rm bins}$ 
index bins in $S$.

The Poisson contribution to the CMB power spectrum can be written as,
\begin{align}
  C_{\ell}(\nu) &= \sum_{i=1}^{N_{\rm sources}} \int_{0}^{\scut}
  dS\, S^2\, P(S_{\nu}|\sobs_{i})
  \notag\\
  &= \sum_{i=1}^{N_{\rm sources}}
  \notag\\
  &\times 
  \sqrt{\frac{\pi}{2}} \frac{\sigma_{S_{\nu}}}{A_{S_{\nu}}}
  \exp\left[\frac{1}{2} (\alpha_{S_{\nu}}+3)((\alpha_{S_{\nu}}+3) \sigma_{S_{\nu}}^2 + 2\mu_{S_{\nu}})\right]
  \notag\\
  &\quad\times
  \left(1 - {\rm Erf}\left[\frac{(\alpha_{S_{\nu}}+3) \sigma_{S_{\nu}}^2 + \mu_{S_{\nu}} - \ln(\scut)}
  {\sqrt{2}\sigma_{S_{\nu}}^2}\right]\right),
\end{align}
where the index $i$ in the final equality is implicit in all the flux posterior parameters 
and ``Erf'' denotes the error function.

\section{NOAO DFS Catalgoue}
\label{sec:catalogtable}
A random sub-sample of the catalog of 
362 sources in the NOAO DFS matched at 1.4 and 5~GHz is shown in 
Table~\ref{tab:noaocat}. The spectral index values in the final column are derived from 
Columns 3 and 5 via the relation $\alpha_{1.4-5}\equiv \log(S_5/S_{1.4}) / \log(5/1.4)$. 
The full catalog is available for download from the VizieR database.
\begin{deluxetable}{lllllll}
\tabletypesize{\scriptsize}
  \tablecaption{\scriptsize Random Sub-sample of the Catalogue of Matched Sources in the NOAO DFS.\label{tab:noaocat}}
  \tablewidth{0.5\textwidth}
  \tablehead{
  \colhead{R.A.} & \colhead{Decl.} & \colhead{$S_{5}$} & \colhead{$\sigma(S_{5})$} & 
  \colhead{$S_{1.4}$} & \colhead{$\sigma(S_{1.4})$} & \colhead{$\alpha_{1.4-5}$}\\
  \colhead{(deg.)} & \colhead{(deg.)} & \colhead{(mJy)} & \colhead{(mJy)} &
  \colhead{(mJy)} & \colhead{(mJy)} & \colhead{}
  }
  \startdata
  34.01473 & -5.13245 & 6.01 & 0.26 & 20.43 & 0.15 & -0.96 \\ 
  30.99103 & -4.96195 & 3.62 & 0.29 & 7.91 & 0.14 & -0.61 \\ 
  30.86276 & -4.98161 & 3.90 & 0.27 & 7.30 & 0.15 & -0.49 \\ 
  33.28917 & -4.09886 & 0.98 & 0.35 & 6.11 & 0.14 & -1.44 \\ 
  33.34937 & -4.33320 & 13.54 & 0.38 & 65.49 & 0.17 & -1.24 \\ 
  33.88087 & -4.68231 & 3.92 & 0.46 & 7.18 & 0.15 & -0.48 \\ 
  31.76899 & -3.81961 & 10.64 & 0.30 & 34.55 & 0.15 & -0.93 \\ 
  33.44730 & -3.71314 & 10.47 & 0.30 & 33.00 & 0.14 & -0.90 \\ 
  33.56511 & -5.52509 & 3.39 & 0.58 & 4.84 & 0.15 & -0.28 \\ 
  31.06313 & -4.72039 & 11.24 & 0.30 & 30.47 & 0.17 & -0.78 \\ 
  33.19339 & -3.61151 & 1.97 & 0.49 & 18.54 & 0.15 & -1.76 \\ 
  32.27030 & -5.10399 & 1.07 & 0.28 & 4.04 & 0.14 & -1.04 \\ 
  31.32263 & -5.27310 & 1.69 & 0.38 & 2.24 & 0.16 & -0.22 \\ 
  32.76205 & -4.89393 & 4.67 & 0.36 & 2.12 & 0.15 & 0.62 \\ 
  34.04510 & -4.43530 & 21.66 & 0.28 & 23.80 & 0.15 & -0.07 \\ 
  \enddata
\end{deluxetable}

\bibliographystyle{apj}
\bibliography{library}

\begin{thebibliography}{39}
\expandafter\ifx\csname natexlab\endcsname\relax\def\natexlab#1{#1}\fi

\bibitem[{{AMI Consortium} {et~al.}(2011){AMI Consortium}, {Davies}, {Franzen},
  {Waldram}, {Grainge}, {Hobson}, {Hurley-Walker}, {Lasenby}, {Olamaie},
  {Pooley}, {Riley}, {Rodr{\'{\i}}guez-Gonz{\'a}lvez}, {Saunders}, {Scaife},
  {Schammel}, {Scott}, {Shimwell}, {Titterington}, \&
  {Zwart}}]{2011MNRAS.415.2708A}
{AMI Consortium}, {Davies}, M.~L., {Franzen}, T.~M.~O., {Waldram}, E.~M.,
  {Grainge}, K.~J.~B., {Hobson}, M.~P., {Hurley-Walker}, N., {Lasenby}, A.,
  {Olamaie}, M., {Pooley}, G.~G., {Riley}, J.~M.,
  {Rodr{\'{\i}}guez-Gonz{\'a}lvez}, C., {Saunders}, R.~D.~E., {Scaife},
  A.~M.~M., {Schammel}, M.~P., {Scott}, P.~F., {Shimwell}, T.~W.,
  {Titterington}, D.~J., \& {Zwart}, J.~T.~L. 2011, \mnras, 415, 2708

\bibitem[{Colombo \& Pierpaoli(2010)}]{2010MNRAS.407..247C}
Colombo, L. P.~L. \& Pierpaoli, E. 2010, \mnras, 407, 247

\bibitem[{{Curto} {et~al.}(2011){Curto}, {Mart{\'{\i}}nez-Gonz{\'a}lez},
  {Barreiro}, \& {Hobson}}]{2011MNRAS.417..488C}
{Curto}, A., {Mart{\'{\i}}nez-Gonz{\'a}lez}, E., {Barreiro}, R.~B., \&
  {Hobson}, M.~P. 2011, \mnras, 417, 488

\bibitem[{Danese {et~al.}(1987)Danese, Franceschini, Toffolatti, \&
  de~Zotti}]{1987ApJ...318L..15D}
Danese, L., Franceschini, A., Toffolatti, L., \& de~Zotti, G. 1987, \apj, 318,
  L15

\bibitem[{de~Zotti {et~al.}(2010)de~Zotti, Massardi, Negrello, \&
  Wall}]{2010A&ARv..18....1D}
de~Zotti, G., Massardi, M., Negrello, M., \& Wall, J. 2010, The \aap Review,
  18, 1

\bibitem[{de~Zotti {et~al.}(2005)de~Zotti, Ricci, Mesa, Silva, Mazzotta,
  Toffolatti, \& Gonz{\'a}lez-Nuevo}]{2005A&A...431..893D}
de~Zotti, G., Ricci, R., Mesa, D., Silva, L., Mazzotta, P., Toffolatti, L., \&
  Gonz{\'a}lez-Nuevo, J. 2005, \aap, 431, 893

\bibitem[{{Elsner} {et~al.}(2010){Elsner}, {Wandelt}, \&
  {Schneider}}]{2010A&A...513A..59E}
{Elsner}, F., {Wandelt}, B.~D., \& {Schneider}, M.~D. 2010, \aap, 513, A59

\bibitem[{Guerra {et~al.}(2002)Guerra, Newlander, Haarsma, \&
  Bruce~Partridge}]{Guerra:2002im}
Guerra, E.~J., Newlander, S.~M., Haarsma, D.~B., \& Bruce~Partridge, R. 2002,
  New Astron. Rev., 46, 303

\bibitem[{Hogg(2008)}]{2008arXiv0807.4820H}
Hogg, D.~W. 2008, arXiv.org, 0807, 4820

\bibitem[{{Huffenberger} {et~al.}(2008){Huffenberger}, {Eriksen}, {Hansen},
  {Banday}, \& {G{\'o}rski}}]{2008ApJ...688....1H}
{Huffenberger}, K.~M., {Eriksen}, H.~K., {Hansen}, F.~K., {Banday}, A.~J., \&
  {G{\'o}rski}, K.~M. 2008, \apj, 688, 1

\bibitem[{Huterer \& Takada(2005)}]{2005APh....23..369H}
Huterer, D. \& Takada, M. 2005, Astropart. Phys., 23, 369

\bibitem[{Knox(1999)}]{1999MNRAS.307..977K}
Knox, L. 1999, \mnras, 307, 977

\bibitem[{Knox {et~al.}(2004)Knox, Holder, \& Church}]{2004ApJ...612...96K}
Knox, L., Holder, G.~P., \& Church, S.~E. 2004, \apj, 612, 96

\bibitem[{Lin {et~al.}(2009)Lin, Partridge, Pober, Bouchefry, Burke, Klein,
  Coish, \& Huffenberger}]{2009ApJ...694..992L}
Lin, Y.-T., Partridge, B., Pober, J.~C., Bouchefry, K.~E., Burke, S., Klein,
  J.~N., Coish, J.~W., \& Huffenberger, K.~M. 2009, \apj, 694, 992

\bibitem[{{Marriage}(2011)}]{2011ApJ...731..100M}
{Marriage}, T.~A. e.~a. 2011, \apj, 731, 100

\bibitem[{Mason {et~al.}(2003)Mason, Pearson, Readhead, Shepherd, Sievers,
  Udomprasert, Cartwright, Farmer, Padin, Myers, Bond, Contaldi, Pen, Prunet,
  Pogosyan, Carlstrom, Kovac, Leitch, Pryke, Halverson, Holzapfel, Altamirano,
  Bronfman, Casassus, May, \& Joy}]{2003ApJ...591..540M}
Mason, B.~S., Pearson, T.~J., Readhead, A. C.~S., Shepherd, M.~C., Sievers, J.,
  Udomprasert, P.~S., Cartwright, J.~K., Farmer, A.~J., Padin, S., Myers,
  S.~T., Bond, J.~R., Contaldi, C.~R., Pen, U., Prunet, S., Pogosyan, D.,
  Carlstrom, J.~E., Kovac, J., Leitch, E.~M., Pryke, C., Halverson, N.~W.,
  Holzapfel, W.~L., Altamirano, P., Bronfman, L., Casassus, S., May, J., \&
  Joy, M. 2003, \apj, 591, 540

\bibitem[{Mason {et~al.}(2009)Mason, Weintraub, Sievers, Bond, Myers, Pearson,
  Readhead, \& Shepherd}]{2009ApJ...704.1433M}
Mason, B.~S., Weintraub, L., Sievers, J., Bond, J.~R., Myers, S.~T., Pearson,
  T.~J., Readhead, A. C.~S., \& Shepherd, M.~C. 2009, \apj, 704, 1433

\bibitem[{Millea {et~al.}(2011)Millea, Dor{\'e}, Dudley, Holder, Knox, Shaw,
  Song, \& Zahn}]{2011arXiv1102.5195M}
Millea, M., Dor{\'e}, O., Dudley, J., Holder, G., Knox, L., Shaw, L., Song,
  Y.-S., \& Zahn, O. 2011, arXiv, 1102, 5195

\bibitem[{Muchovej {et~al.}(2010)Muchovej, Leitch, Carlstrom, Culverhouse,
  Greer, Hawkins, Hennessy, Joy, Lamb, Loh, Marrone, Miller, Mroczkowski,
  Pryke, Sharp, \& Woody}]{2010ApJ...716..521M}
Muchovej, S., Leitch, E., Carlstrom, J.~E., Culverhouse, T., Greer, C.,
  Hawkins, D., Hennessy, R., Joy, M., Lamb, J., Loh, M., Marrone, D.~P.,
  Miller, A., Mroczkowski, T., Pryke, C., Sharp, M., \& Woody, D. 2010, \apj,
  716, 521

\bibitem[{{Paoletti} {et~al.}(2011){Paoletti}, {Aghanim}, {Douspis}, {Finelli},
  {De Zotti}, {Lagache}, \& {P{\'e}nin}}]{2011arXiv1112.3260P}
{Paoletti}, D., {Aghanim}, N., {Douspis}, M., {Finelli}, F., {De Zotti}, G.,
  {Lagache}, G., \& {P{\'e}nin}, A. 2011, arXiv:1112.3260

\bibitem[{{Perna} \& {Di Matteo}(2000)}]{2000ApJ...542...68P}
{Perna}, R. \& {Di Matteo}, T. 2000, \apj, 542, 68

\bibitem[{{Pierpaoli} \& {Perna}(2004)}]{2004MNRAS.354.1005P}
{Pierpaoli}, E. \& {Perna}, R. 2004, \mnras, 354, 1005

\bibitem[{{Planck Collaboration} {et~al.}(2011){Planck Collaboration}, {Ade},
  {Aghanim}, {Arnaud}, {Ashdown}, {Aumont}, {Baccigalupi}, {Balbi}, {Banday},
  {Barreiro}, \& et~al.}]{PlanckERCSC}
{Planck Collaboration}, {Ade}, P.~A.~R., {Aghanim}, N., {Arnaud}, M.,
  {Ashdown}, M., {Aumont}, J., {Baccigalupi}, C., {Balbi}, A., {Banday}, A.~J.,
  {Barreiro}, R.~B., \& et~al. 2011, \aap, 536, A7

\bibitem[{Prandoni {et~al.}(2010)Prandoni, de~Ruiter, Ricci, Parma, Gregorini,
  \& Ekers}]{Prandoni:2010gv}
Prandoni, I., de~Ruiter, H.~R., Ricci, R., Parma, P., Gregorini, L., \& Ekers,
  R.~D. 2010, \aap, 510, A42

\bibitem[{{Prandoni} {et~al.}(2000){Prandoni}, {Gregorini}, {Parma}, {de
  Ruiter}, {Vettolani}, {Wieringa}, \& {Ekers}}]{2000A&AS..146...41P}
{Prandoni}, I., {Gregorini}, L., {Parma}, P., {de Ruiter}, H.~R., {Vettolani},
  G., {Wieringa}, M.~H., \& {Ekers}, R.~D. 2000, \aaps, 146, 41

\bibitem[{Prandoni {et~al.}(2006)Prandoni, Parma, Wieringa, de~Ruiter,
  Gregorini, Mignano, Vettolani, \& Ekers}]{Prandoni:2006rm}
Prandoni, I., Parma, P., Wieringa, M., de~Ruiter, H.~R., Gregorini, L.,
  Mignano, A., Vettolani, G., \& Ekers, R.~D. 2006, \aap, 457, 517

\bibitem[{{Rani} {et~al.}(2011){Rani}, {Gupta}, {Bachev}, {Strigachev},
  {Semkov}, {D'Ammando}, {Wiita}, {Gurwell}, {Ovcharov}, {Mihov}, {Boeva}, \&
  {Peneva}}]{2011MNRAS.417.1881R}
{Rani}, B., {Gupta}, A.~C., {Bachev}, R., {Strigachev}, A., {Semkov}, E.,
  {D'Ammando}, F., {Wiita}, P.~J., {Gurwell}, M.~A., {Ovcharov}, E., {Mihov},
  B., {Boeva}, S., \& {Peneva}, S. 2011, \mnras, 417, 1881

\bibitem[{{Reese} {et~al.}(2011){Reese}, {Mroczkowski}, {Menanteau}, {Hilton},
  {Sievers}, {Aguirre}, {Appel}, {Baker}, {Bond}, {Das}, {Devlin}, {Dicker},
  {Dunner}, {Essinger-Hileman}, {Fowler}, {Hajian}, {Halpern}, {Hasselfield},
  {Hill}, {Hincks}, {Huffenberger}, {Hughes}, {Irwin}, {Klein}, {Kosowsky},
  {Lin}, {Marriage}, {Marsden}, {Moodley}, {Niemack}, {Nolta}, {Page},
  {Parker}, {Partridge}, {Rojas}, {Sehgal}, {Sifon}, {Spergel}, {Staggs},
  {Swetz}, {Switzer}, {Thornton}, {Trac}, \& {Wollack}}]{2011arXiv1108.3343R}
{Reese}, E.~D., {Mroczkowski}, T., {Menanteau}, F., {Hilton}, M., {Sievers},
  J., {Aguirre}, P., {Appel}, J.~W., {Baker}, A.~J., {Bond}, J.~R., {Das}, S.,
  {Devlin}, M.~J., {Dicker}, S.~R., {Dunner}, R., {Essinger-Hileman}, T.,
  {Fowler}, J.~W., {Hajian}, A., {Halpern}, M., {Hasselfield}, M., {Hill},
  J.~C., {Hincks}, A.~D., {Huffenberger}, K.~M., {Hughes}, J.~P., {Irwin},
  K.~D., {Klein}, J., {Kosowsky}, A., {Lin}, Y.-T., {Marriage}, T.~A.,
  {Marsden}, D., {Moodley}, K., {Niemack}, M.~D., {Nolta}, M.~R., {Page},
  L.~A., {Parker}, L., {Partridge}, B., {Rojas}, F., {Sehgal}, N., {Sifon}, C.,
  {Spergel}, D.~N., {Staggs}, S.~T., {Swetz}, D.~S., {Switzer}, E.~R.,
  {Thornton}, R., {Trac}, H., \& {Wollack}, E.~J. 2011, arXiv:1108.3343

\bibitem[{Schneider {et~al.}(2011)Schneider, Holm, \&
  Knox}]{2011ApJ...728..137S}
Schneider, M.~D., Holm, O., \& Knox, L. 2011, \apj, 728, 137

\bibitem[{Scott \& White(1999)}]{1999AA...346....1S}
Scott, D. \& White, M. 1999, \aap, 346, 1

\bibitem[{Sehgal {et~al.}(2010)Sehgal, Bode, Das, Hernandez-Monteagudo,
  Huffenberger, Lin, Ostriker, \& Trac}]{2010ApJ...709..920S}
Sehgal, N., Bode, P., Das, S., Hernandez-Monteagudo, C., Huffenberger, K., Lin,
  Y.-T., Ostriker, J.~P., \& Trac, H. 2010, \apj, 709, 920

\bibitem[{Tegmark \& Efstathiou(1996)}]{1996MNRAS.281.1297T}
Tegmark, M. \& Efstathiou, G. 1996, \mnras, 281, 1297

\bibitem[{Toffolatti {et~al.}(1998)Toffolatti, Argueso~Gomez, de~Zotti, Mazzei,
  Franceschini, Danese, \& Burigana}]{1998MNRAS.297..117T}
Toffolatti, L., Argueso~Gomez, F., de~Zotti, G., Mazzei, P., Franceschini, A.,
  Danese, L., \& Burigana, C. 1998, \mnras, 297, 117

\bibitem[{{Toffolatti} {et~al.}(2005){Toffolatti}, {Negrello},
  {Gonz{\'a}lez-Nuevo}, {de Zotti}, {Silva}, {Granato}, \&
  {Arg{\"u}eso}}]{2005A&A...438..475T}
{Toffolatti}, L., {Negrello}, M., {Gonz{\'a}lez-Nuevo}, J., {de Zotti}, G.,
  {Silva}, L., {Granato}, G.~L., \& {Arg{\"u}eso}, F. 2005, \aap, 438, 475

\bibitem[{{Tucci} {et~al.}(2011){Tucci}, {Toffolatti}, {de Zotti}, \&
  {Mart{\'{\i}}nez-Gonz{\'a}lez}}]{2011A&A...533A..57T}
{Tucci}, M., {Toffolatti}, L., {de Zotti}, G., \&
  {Mart{\'{\i}}nez-Gonz{\'a}lez}, E. 2011, \aap, 533, A57

\bibitem[{Vieira {et~al.}(2010)Vieira, Crawford, Switzer, Ade, Aird, Ashby,
  Benson, Bleem, Brodwin, Carlstrom, Chang, Cho, Crites, de~Haan, Dobbs,
  Everett, George, Gladders, Hall, Halverson, High, Holder, Holzapfel, Hrubes,
  Joy, Keisler, Knox, Lee, Leitch, Lueker, Marrone, McIntyre, McMahon, Mehl,
  Meyer, Mohr, Montroy, Padin, Plagge, Pryke, Reichardt, Ruhl, Schaffer, Shaw,
  Shirokoff, Spieler, Stalder, Staniszewski, Stark, Vanderlinde, Walsh,
  Williamson, Yang, Zahn, \& Zenteno}]{2010ApJ...719..763V}
Vieira, J.~D., Crawford, T.~M., Switzer, E.~R., Ade, P. A.~R., Aird, K.~A.,
  Ashby, M. L.~N., Benson, B.~A., Bleem, L.~E., Brodwin, M., Carlstrom, J.~E.,
  Chang, C.~L., Cho, H.~M., Crites, A.~T., de~Haan, T., Dobbs, M.~A., Everett,
  W., George, E.~M., Gladders, M., Hall, N.~R., Halverson, N.~W., High, F.~W.,
  Holder, G.~P., Holzapfel, W.~L., Hrubes, J.~D., Joy, M., Keisler, R., Knox,
  L., Lee, A.~T., Leitch, E.~M., Lueker, M., Marrone, D.~P., McIntyre, V.,
  McMahon, J.~J., Mehl, J., Meyer, S.~S., Mohr, J.~J., Montroy, T.~E., Padin,
  S., Plagge, T., Pryke, C., Reichardt, C.~L., Ruhl, J.~E., Schaffer, K.~K.,
  Shaw, L., Shirokoff, E., Spieler, H.~G., Stalder, B., Staniszewski, Z.,
  Stark, A.~A., Vanderlinde, K., Walsh, W., Williamson, R., Yang, Y., Zahn, O.,
  \& Zenteno, A. 2010, \apj, 719, 763

\bibitem[{White(1998)}]{1998PhRvD..57.5273W}
White, M. 1998, \prd (Particles, 57, 5273

\bibitem[{White \& Majumdar(2004)}]{2004ApJ...602..565W}
White, M. \& Majumdar, S. 2004, \apj, 602, 565

\bibitem[{White {et~al.}(1997)White, Becker, Helfand, \&
  Gregg}]{1997ApJ...475..479W}
White, R.~L., Becker, R.~H., Helfand, D.~J., \& Gregg, M.~D. 1997, \apj, 475,
  479

\end{thebibliography}

\end{document}